\documentclass[journal,draftclsnofoot,onecolumn,12pt,twoside]{IEEEtran}
\usepackage[T1]{fontenc}
\usepackage[latin9]{inputenc}
\usepackage{verbatim}
\usepackage{float}
\usepackage{textcomp}
\usepackage{amsmath}
\usepackage{amsthm}
\usepackage{amssymb}
\usepackage{graphicx}
\usepackage[unicode=true,
bookmarks=true,bookmarksnumbered=true,bookmarksopen=true,bookmarksopenlevel=1,
breaklinks=false,pdfborder={0 0 0},pdfborderstyle={},backref=false,colorlinks=false]
{hyperref}
\hypersetup{pdftitle={Your Title},
	pdfauthor={Your Name},
	pdfpagelayout=OneColumn, pdfnewwindow=true, pdfstartview=XYZ, plainpages=false}

\makeatletter

\providecommand{\tabularnewline}{\\}
\floatstyle{ruled}
\newfloat{algorithm}{tbp}{loa}
\providecommand{\algorithmname}{Algorithm}
\floatname{algorithm}{\protect\algorithmname}

\theoremstyle{plain}
\newtheorem{thm}{\protect\theoremname}
\theoremstyle{remark}
\newtheorem{rem}[thm]{\protect\remarkname}
\theoremstyle{plain}
\newtheorem{lem}[thm]{\protect\lemmaname}

\usepackage{algorithmic}
\usepackage{cite}
\usepackage{siunitx}
\floatname{algorithm}{Algorithm}

\DeclareMathOperator{\maximize}{maximize}
\DeclareMathOperator{\minimize}{minimize}
\DeclareMathOperator{\st}{subject \, to}

\newcommand{\herm}{^{{\dagger}}}
\newcommand{\trans}{^{\mbox{\scriptsize T}}}
\usepackage{subfig}

\@ifundefined{showcaptionsetup}{}{%
	\PassOptionsToPackage{caption=false}{subfig}}
\usepackage{subfig}
\makeatother

\providecommand{\lemmaname}{Lemma}
\providecommand{\remarkname}{Remark}
\providecommand{\theoremname}{Theorem}

\begin{document}
\title{Utility Maximization for Large-Scale Cell-Free Massive MIMO Downlink}
\author{Muhammad~Farooq, \IEEEmembership{Student Member, IEEE}, Hien~Quoc~Ngo, \IEEEmembership{Senior Member, IEEE}, Een-Kee Hong, \IEEEmembership{Senior Member, IEEE},	and~Le-Nam~Tran, \IEEEmembership{Senior Member, IEEE}
	\thanks{Muhammad~Farooq and Le-Nam~Tran are with the School of Electrical
		and Electronic Engineering, University College Dublin, Ireland (e-mail: muhammad.farooq@ucdconnect.ie;nam.tran@ucd.ie).}
	\thanks{Hien~Quoc~Ngo is with the Institute of Electronics, Communications
		and Information Technology, Queen\textquoteright s University Belfast,
		Belfast BT3 9DT, U.K. (email: hien.ngo@qub.ac.uk).}
	\thanks{Een-Kee Hong is with the Department of Electronic Engineering Kyung Hee University, Yong-in, Republic of Korea. (email: ekhong@khu.ac.kr).}
	\thanks{Parts of this work will be presented at IEEE PIMRC 2020 \cite{Farooq2020}.}}%

\maketitle

\begin{abstract}
We consider the system-wide utility maximization problem in the downlink of a cell-free massive multiple-input multiple-output (MIMO) system whereby \emph{a very large number} of access points (APs) simultaneously serve a group of users. Specifically, four fundamental problems with increasing order of user fairness are of interest: (i) to maximize the average spectral efficiency (SE), (ii) to maximize the proportional fairness, (iii) to maximize the harmonic-rate of all users, and lastly (iv) to maximize the minimum SE of all users, subject to a sum power constraint at each AP. As the considered problems are non-convex, existing solutions normally rely on successive convex approximation to find a sub-optimal solution. More specifically, these known methods use off-the-shelf convex solvers, which basically implement an interior-point algorithm, to solve the derived convex problems. The main issue of such methods is that their complexity does not scale favorably with the problem size, limiting previous studies to cell-free massive MIMO of moderate scales. Thus the potential of cell-free massive MIMO has not been fully understood. To address this issue, we propose a unified framework based on an accelerated projected gradient method to solve the considered problems. Particularly, the proposed solution is found in closed-form expressions and only requires the first order oracle of the objective, rather than the Hessian matrix as in known solutions, and thus is much more memory efficient. Numerical results demonstrate that our proposed solution achieves the same utility performance but with far less run-time, compared to other second-order methods. Simulation results for large-scale cell-free massive MIMO show that the four utility functions can deliver nearly uniformed services to all users. In other words, user fairness is not a great concern in large-scale cell-free massive MIMO.
\end{abstract}


\IEEEpeerreviewmaketitle

\section{Introduction}

\IEEEPARstart{M}{ultiple}-input multiple-output (MIMO) is the underlying technology
in the physical layer of many modern wireless communications standards.
The use of multiple antennas at transceivers can offer high data rates
and high reliability by exploiting spatial and diversity gains \cite{Telatar:MIMO:1999,Foschini:MIMO:1998,Alamouti:STBC:1998}.
To meet a set of requirements for 5G networks, MIMO has evolved into
so-called massive MIMO where a very large number of antennas are deployed
at each base station (BS) to serve many users at the same time \cite{Mazretta2010,Boccardi2014}. In particular,
massive MIMO has been implemented in the first version of 5G NR \cite{Dahlman2018}. Since 5G still follows the conventional design of a \emph{cellular}
network like its predecessors, inter-cell interference remains a fundamental
problem, and thus massive MIMO cannot be unlocked to its full potential
\cite{Lozano2013}.

There are two types of massive MIMO in terms of the service area:
colocated massive MIMO and distributed massive MIMO. For the former,
all the antennas are placed in a small area and therefore, the processing
complexity requirement is very low. For the latter, on the other hand,
the antennas are distributed to serve a relatively much larger area.
These systems are more diverse against the shadow fading and they have a large coverage  area \cite{Yang2014}. There is no doubt that
the distributed massive MIMO is better than colocated massive MIMO
but due to the more processing complexity and high cost requirements
\cite{BJORNSON20193}, the scalability remains an active area of research
in distributed systems.

Cell-free massive multiple-input multiple-output (MIMO) was introduced
in \cite{Ngo2017} as a major leap of massive MIMO technology to overcome
the inter-cell interference which is the main inherent limitation
of cellular-based networks. In cell-free massive MIMO, many access
points (APs) distributed over the whole network serve many users in
the same time-frequency resource. There are no cells, and hence, no
boundary effects. Unlike colocated massive MIMO, each AP in cell-free massive MIMO is equipped
with just a few antennas. But an important point is that when the
number of APs is very large, cell-free massive MIMO is still able
to exploit the favorable propagation and channel hardening properties,
like colocated massive MIMO. In particular,  with favorable propagation, APs can use simple linear processing techniques to combine the signals in the uplink, and precode the symbols in the downlink. With channel hardening, decoding the signals using the channel statistics (large-scale fading coefficients) can provide good performance. In addition, all resource allocations (i.e. power control, user scheduling or AP selections) can be done over the large-scale fading time scale \cite{Ngo2018EE,zhang2019cell}. Note that in some propagation environments, the level of channel hardening in cell-free massive MIMO is lesser than that in colocated massive MIMO \cite{chen2018channel}.

The research on cell-free massive MIMO is still in its infancy and
thus deserves more extensive and thorough studies. We discuss here
some of the noticeable and related studies in the literature. In \cite{Ngo2017},
Ngo \emph{et al.} considered the problem of minimum rate maximization
to provide uniformly good services to all users. The problem was then
solved using a bisection search and a sequence of second-order cone feasibility
problems. In \cite{Nguyen2017}, Nguyen \emph{et al.} adopted zero-forcing
precoding and studied the energy efficiency maximization (EEmax) problem.
In this work, an iterative method based on successive convex approximation
(SCA) was derived. In \cite{Buzzi2017}, both the max-min fairness
and sum-rate maximization problems for user-centric cell-free massive MIMO were considered and solved by SCA.
The SCA-based method was also used in \cite{Ngo2018EE} and \cite{Interdonato2019}
to solve the EEmax and max-min fairness power controls with different
cell-free massive MIMO setups, respectively.

A common feature of all the above mentioned pioneer studies on cell-free
massive MIMO is the use of a second-order interior-point method which
requires the computation of the Hessian matrix of the objective, and
thus their computational complexity and memory requirement makes them
impossible to implement and investigate the performance of large-scale cell-free
massive MIMO. To motivate our proposed method, let us consider an
example where \num{2000} APs are deployed to serve \num{200} users
over an area of \SI{1}{\km\squared}, which is typical for an urban
area in our vision. The power control problem arising from this scenario
has $4\times10^{5}$ optimization variables. Consequently, we would
basically need\textbf{ }160 GB of memory to store the resulting Hessian
matrix, assuming a single-precision floating-point format. It is this
immense memory requirement of the existing power control methods that
only allows us to implement as well as to characterize the performance of cell-free massive
MIMO for a relatively small-scale system. For example, the work of \cite{Ngo2017}
was able to consider an area of \SI{1}{\square\km}, consisting of
100 APs serving 40 users. Numbers with the same order of magnitude
were also observed in the above mentioned papers. The performance
of these scenarios fractionally represents the full potential of cell-free
massive MIMO.

To fully understand the performance limits of cell-free massive MIMO and to make the power controls feasible for practical implementation,
we urgently need to devise more memory efficient power control methods.
To this end, we propose in this paper a first order method to maximize
several system-wide utility functions, namely the total spectral efficiency,
proportional fairness, harmonic-rate, and the minimum rate of the
downlink. Referring to the motivating example in the preceding paragraph,
a first order optimization method only requires a memory of 8 MB,
which is affordable by most, if not all, modern desktops. In this
paper, similar to many previous studies (e.g. \cite{Ngo2018EE}),
we adopt the conjugate beamforming at each AP, and the power control
problems for utility maximization are based on large-scale fading. While
first-order methods are popular for convex programming, they are relatively
open for nonconvex programming which is unfortunately the case for
the considered problems. In the paper, we capitalize on the accelerated
proximal gradient method for nonconvex programming presented
in \cite{Li2015} to derive efficient and unified solutions to the
considered utility maximization problems. A similar method has been
used in \cite{Nam2019APG} to solve the EEmax problem. Our contributions
are as follows
\begin{itemize}
	\item We first present a brief introduction of an accelerated proximal gradient
	method in general and a special variant in particular, which we refer
	to as accelerated projected gradient (APG) method, for nonconvex programming.
	\item We propose iterative power control algorithms drawing on the APG method
	to efficiently solve the considered utility maximization problems.
	Particularly, each iteration of the proposed algorithms is done by
	closed-form expressions and can be done in parallel. To achieve this,
	we reformulate the considered problems so that the gradient of the
	objective is Lipchitz continuous and the projection is still efficient
	to compute.
	\item We provide a  complexity and convergence analysis of the proposed
	methods. Specifically, the per-iteration complexity of our proposed
	method is only $\mathcal{O}(K^{2}M)$ as compared to the per-iteration
	complexity of $\mathcal{O}\left(\sqrt{K+M}M^{3}K^{4}\right)$ for
	the SCA-method in \cite{Ngo2018EE}, where $M$ and $K$ are the numbers of APs and users, respectively. Accordingly, the proposed method
	takes much reduced run time to return a solution as numerically shown
	in Section \ref{sec:Numerical-Results}. As a result, the propose
	method can lay the foundation to numerically analyze the performance
	of large-scale cell-free massive MIMO.
	\item We carry out extensive numerical experiments to draw useful insights
	into the performance of large-scale cell-free massive MIMO regarding
	the four utility metrics above. In particular we find that, in the
	domain of large-scale cell-free massive MIMO, per-user rates are quite
	comparable for the four above utility functions, which means that
	large-scale cell-free massive MIMO can deliver universally good services
	to all users. Also, in terms of per-user rate, it is more beneficial to use a higher number
	of APs with a fewer antenna per AP than to use a smaller number of APs
	with more antennas per APs.
\end{itemize}
\emph{Notation}s: Bold lower and upper case letters represent vectors
and matrices. $\mathcal{CN}(0,a)$ denotes a complex Gaussian random
variable with zero mean and variance $a$. $\mathbf{X}\trans$ and
$\mathbf{X}\herm$ stand for the transpose and Hermitian of $\mathbf{X}$,
respectively. $x_{i}$ is the $i$-th entry of vector $\mathbf{x}$;
$[\mathbf{X}]_{i,j}$ is the $(i,j)$-th entry of $\mathbf{X}$. $\nabla f(\mathbf{x})$
represents the gradient of $f(\mathbf{x})$ and $\frac{\partial}{\partial\mathbf{x}_{i}}f(\mathbf{x})$
is the partial gradient with respect to $\mathbf{x}_{i}$. $\left\langle \mathbf{x},\mathbf{y}\right\rangle \triangleq\mathbf{x}\trans\mathbf{y}$
is the inner product of vectors $\mathbf{x}$ and $\mathbf{y}$. $[\mathbf{x}]_{+}$
denotes the projector onto the positive orthant. $||\cdot||$ represents
the Euclidean norm; $|\cdot|$ is the absolute value of the argument.

\section{System Model and Problem Formulation}

\subsection{System Model}

We consider the downlink of a cell-free massive MIMO system model
as in \cite{Ngo2018EE}. In particular, there are $M$ APs serving
$K$ single-antenna users in time division duplex (TDD) mode. Each
AP is equipped with $N$ antennas. All the APs and the users are assumed
to be distributed in a large area. As TDD operation is adopted, APs
first estimate the channels using pilot sequences from the uplink
(commonly known as uplink training) and then apply a beamforming technique
to transmit signals to all users in the downlink, or use a matched
filter technique to combine signals in the uplink. Since this work
focuses on the downlink transmission, we neglect the uplink payload
transmission phase. Let us denote by $T_{c}$ and $T_{p}$ the length
of the coherence interval and the uplink training phase in data symbols, respectively.
The uplink training and downlink payload transmission phases are summarized
as follows. The interested reader is referred to \cite{Ngo2018EE}
for further details.

\subsubsection{Uplink Training}

We assume the channel is reciprocal, i.e., the channel gains on the
uplink and on the downlink are the same. Consequently, APs can estimate
the downlink channel based on the pilot sequences sent by all users
on the uplink. Let $\sqrt{T_{p}}\boldsymbol{\psi}_{k}\in\mathbb{C}^{T_{p}\times1}$,
where $||\boldsymbol{\psi}_{k}||^{2}=1$, be the pilot sequence transmitted
from the $k$-th user, $k=1,\ldots,K$. Note that $T_{p}$ is the
length of the pilot sequences, which is the same for all users. 
%These pilot sequences are assumed to be orthonormal and thus the effectof pilot contamination is ignored. 
The received signal at the $m$-th
AP is given by 
\begin{equation}
\mathbf{R}_{\textrm{up},m}=\sqrt{\zeta_{p}T_{p}}\sum_{k=1}^{K}\mathbf{g}_{mk}\boldsymbol{\psi}_{k}\herm+\mathbf{W}_{\textrm{up},m},
\end{equation}
where $\zeta_{p}$ is the normalized transmit signal-to-noise ratio
(SNR) of each pilot symbol, and $\mathbf{W}_{\textrm{up},m}\in\mathbb{C}^{N\times T_{p}}$
is the noise matrix whose entries are independent and identically
distributed (i.i.d.) drawn from $\mathcal{CN}(0,1)$, and $\mathbf{g}_{mk}\in\mathbb{C}^{N\times1}$
is the channel between the $m$-th AP and the $k$-th user. As in
\cite{Ngo2018EE}, we model $\mathbf{g}_{mk}$ as 
\begin{equation}
\mathbf{g}_{mk}=\beta_{mk}^{1/2}\mathbf{h}_{mk},
\end{equation}
where $\text{\ensuremath{\beta_{mk}}}$ represents the large-scale
fading (i.e. including path loss and shadowing effects) and $\mathbf{h}_{mk}\in\mathbb{C}^{N\times1}$
comprises of small-scale fading coefficients between the $N$ antennas
of the $m$-th AP and the $k$-th user. We further assume that the
entries of $\mathbf{h}_{mk}$ follows i.i.d. $\mathcal{CN}(0,1)$.

Next the $m$-th AP needs to estimate the channel $\mathbf{g}_{mk}$,
$k=1,2,\ldots,N$, based on the received pilot signal $\mathbf{R}_{\textrm{up},m}$.
To do so, the $m$-th AP projects $\mathbf{R}_{\textrm{up},m}$ onto
$\boldsymbol{\psi}_{k}$, producing 
\begin{equation}
\mathbf{r}_{mk}=\mathbf{R}_{\textrm{up},m}\boldsymbol{\psi}_{k}=\sqrt{\zeta_{p}T_{p}}\sum_{i=1}^{K}\mathbf{g}_{mi}\boldsymbol{\psi}_{i}\herm\boldsymbol{\psi}_{k}+\tilde{\mathbf{w}}_{mk},
\end{equation}
where $\tilde{\mathbf{w}}_{mk}\triangleq\mathbf{W}_{\textrm{up},m}\boldsymbol{\psi}_{k}\in\mathbb{C}^{N\times1}$
has entries following i.i.d. $\mathcal{CN}(0,1)$. %As explained in \cite{Ngo2017}, $\mathbf{r}_{mk}$ is a sufficient statistic for orthonormal pilot sequences. 
Given $\mathbf{r}_{mk}$, the minimum mean-square error (MMSE)
of the channel estimate of ${\mathbf{g}}_{mk}$ is \cite{Ngo2018EE}
\begin{equation}
\hat{\mathbf{g}}_{mk} =\mathbb{E}\{\mathbf{g}_{mk}\mathbf{r}_{mk}\herm\}\Bigl(\mathbb{E}\{\mathbf{r}_{mk}\mathbf{r}_{mk}\herm\}\Bigr)^{-1}\mathbf{r}_{mk} =\frac{\sqrt{\zeta_{p}T_{p}}\beta_{mk}}{1+\zeta_{p}T_{p}\sum_{i=1}^{K}\beta_{mi}\left|\boldsymbol{\psi}_{i}\herm\boldsymbol{\psi}_{k}\right|^{2}}\mathbf{r}_{mk}.\label{eq:channel_estimate}
\end{equation}
Note that the expectations in the above equation are carried out with
respect to small-scale fading and also that elements of $\hat{\mathbf{g}}_{mk}$
are independent and identical Gaussian distribution. The mean square
of any element of $\hat{\mathbf{g}}_{mk}$ is given by
\begin{equation}
\nu_{mk}=\frac{\zeta_{p}T_{p}\beta_{mk}^{2}}{1+\zeta_{p}T_{p}\sum_{i=1}^{K}\beta_{mi}\left|\boldsymbol{\psi}_{i}\herm\boldsymbol{\psi}_{k}\right|^{2}}.
\end{equation}

\subsubsection{Downlink Payload Data Transmission}

For downlink payload data transmission, the APs use the channel estimates
obtained in \eqref{eq:channel_estimate} to form separate radio beams
to the $K$ users. As mentioned earlier we adopt conjugate beamforming
in this paper, which is due to two main reasons. First, conjugate
beamforming is computationally simple and can be done locally at each
AP. Second, conjugate beamforming offers excellent performance for
a large number of APs (relatively compared to the number of users).
Denote the symbol to be sent to the $k$-th user by $c_{k}$ and the
power control coefficient between the $m$-th AP and the $k$-th user
by $\eta_{mk}$. For conjugate beamforming, the transmitted symbols
from the $m$-th AP are contained in the vector $\mathbf{x}_{m}$
given by 
\begin{equation}
\mathbf{x}_{m}=\sqrt{\zeta_{d}}\sum_{k=1}^{K}\sqrt{\eta_{mk}}\hat{\mathbf{g}}_{mk}^{*}c_{k},\label{eq:transmitvector}
\end{equation}
where $\zeta_{d}$ is the maximum downlink transmit power at each
AP normalized to noise power. Note that the total power at each AP
is 
\begin{equation}
\mathbb{E}\{||\mathbf{x}_{m}||^{2}\}=\zeta_{d}N\sum_{k=1}^{K}\eta_{mk}\nu_{mk}.\label{eq:spc:origin}
\end{equation}
The received signal at the $k$-th user is written as 
\begin{align}
r_{k} & =\sum_{m=1}^{M}\mathbf{g}_{mk}\trans\mathbf{x}_{m}+w_{k}\nonumber \\
& =\sqrt{\zeta_{d}}\sum_{m=1}^{M}\sqrt{\eta_{mk}}\mathbf{g}_{mk}\trans\hat{\mathbf{g}}_{mk}^{*}c_{k}
+\sqrt{\zeta_{d}}\sum_{i\neq k}^{K}\sum_{m=1}^{M}\sqrt{\eta_{mi}}\mathbf{g}_{mk}\trans\hat{\mathbf{g}}_{mi}^{*}c_{i}+w_{k},\label{eq:receivedsig}
\end{align}
where $w_{k}$ is the white Gaussian noise with zero mean and unit
variance.

\subsubsection{Signal Detection based on Channel Statistics and Spectral Efficiency}

Ideally, to detect $c_{k}$, the $k$-th user needs to know the effective
channel gain $\sqrt{\zeta_{d}}\sum_{m=1}^{M}\sqrt{\eta_{mk}}\mathbf{g}_{mk}\trans\hat{\mathbf{g}}_{mk}^{*}$.
However this is impossible since there are no downlink pilots. Instead,
the $k$-th user will rely on the mean of the effective channel gain
to detect $c_{k}$. To see this we rewrite \eqref{eq:receivedsig}
as 
\begin{align}
r_{k} & =\sqrt{\zeta_{d}}\mathbb{E}\left\{ \sum_{m=1}^{M}\sqrt{\eta_{mk}}\mathbf{g}_{mk}\trans\hat{\mathbf{g}}_{mk}^{*}\right\} c_{k}
+ \sqrt{\zeta_{d}}\left(\sum_{m=1}^{M}\sqrt{\eta_{mk}}\mathbf{g}_{mk}\trans\hat{\mathbf{g}}_{mk}^{*}-\mathbb{E}\left\{ \sum_{m=1}^{M}\sqrt{\eta_{mk}}\mathbf{g}_{mk}\trans\hat{\mathbf{g}}_{mk}^{*}\right\} \right)c_{k}\nonumber \\
& \qquad+\sum_{i\neq k}^{K}\sqrt{\zeta_{d}}\sum_{m=1}^{M}\sqrt{\eta_{mi}}\mathbf{g}_{mk}\trans\hat{\mathbf{g}}_{mi}^{*}c_{i}+w_{k}.
\end{align}
In the above equation the second term in the right side can be seen
as the beamforming uncertainty, which is due to treating the mean
of the effective channel gain as the true channel. We remark that
by the law of large numbers which holds for our system model, with high probability, this
term is much smaller compared to the mean of the effective channel
gain. By further treating this and the inter-user interference as
the Gaussian noise, we can express the signal to interference plus
noise ratio (SINR) at the $k$-th user as\cite[Appendix A]{Ngo2018EE}
\begin{equation}
\gamma_{k}(\bar{\boldsymbol{\eta}})=\frac{\zeta_{d}N^{2}\left|\boldsymbol{\nu}_{kk}\trans\bar{\boldsymbol{\eta}}_{k}\right|^{2}}{\zeta_{d}N^{2}\sum_{i\neq k}^{K}\left|\boldsymbol{\nu}_{ik}\trans\bar{\boldsymbol{\eta}}_{i}\right|^{2}+\zeta_{d}N\sum_{i=1}^{K}||\mathbf{D}_{ik}\bar{\boldsymbol{\eta}}_{i}||_{2}^{2}+1},
\end{equation}
where $\bar{\boldsymbol{\eta}}_{k}=[\sqrt{\eta_{1k}};\ldots;\sqrt{\eta_{Mk}}]\in\mathbb{R}_{+}^{M}$
consists of all power control coefficients associated with user $k$,
$\boldsymbol{\bar{\eta}}=[\boldsymbol{\bar{\eta}}_{1};\bar{\boldsymbol{\eta}}_{2};\ldots;\bar{\boldsymbol{\eta}}_{K}]\in\mathbb{R}_{+}^{MK}$,
$\mathbf{D}_{ik}\in\mathbb{R}_{+}^{M\times M}$ is a diagonal matrix
with $[\mathbf{D}_{ik}]_{m,m}=\sqrt{\nu_{mi}\beta_{mk}}$, and 
\begin{equation}
\boldsymbol{\nu}_{ik}\triangleq\left|\boldsymbol{\psi}_{i}\herm\boldsymbol{\psi}_{k}\right|\left[\nu_{1i}\frac{\beta_{1k}}{\beta_{1i}};\nu_{2i}\frac{\beta_{2k}}{\beta_{2i}};\ldots;\nu_{Mi}\frac{\beta_{Mk}}{\beta_{Mi}}\right].
\end{equation}
Accordingly, the spectral efficiency of the $k$-th user is given
by 
\begin{equation}
\mathrm{SE}_{k}(\bar{\boldsymbol{\eta}})=\Bigl(1-\frac{T_{p}}{T_{c}}\Bigr)\log\left(1+\gamma_{k}(\bar{\boldsymbol{\eta}})\right)\ (\textrm{nat/s/Hz}).\label{eq:SEk}
\end{equation}
Note that for mathematical convenience we use the natural logarithm
in \eqref{eq:SEk}, and thus the resulting unit of the SE is nat/s/Hz.
However, for the numerical results presented in Section \ref{sec:Numerical-Results},
we instead use the logarithm to base $2$ to compute the SE and the
corresponding unit is bit/s/Hz.

\subsection{\label{subsec:Problem-Formulation}Problem Formulation}

To formulate the considered problem and to facilitate the development
of the proposed algorithm, we define $\boldsymbol{\mu}_{m}\in\mathbb{R}_{+}^{K}$
to be the vector of all power control coefficients associated with
the $m$-th AP as 
\begin{equation}
\boldsymbol{\mu}_{m}\triangleq[\mu_{m1};\mu_{m2};\ldots;\mu_{mK}],\label{eq:reformulation1}
\end{equation}
where $\mu_{mk}=\sqrt{\eta_{mk}\nu_{mk}},m=1,\ldots,M,k=1,\ldots,K.$
We also define $\boldsymbol{\mu}\triangleq[\boldsymbol{\mu}_{1};\boldsymbol{\mu}_{2};\ldots;\boldsymbol{\mu}_{M}]\in\mathbb{R}_{+}^{MK\times1}$
to include the power control coefficients of all APs. To express the
spectral efficiency in \eqref{eq:SEk} as a function of $\boldsymbol{\mu}$,
we denote by $\bar{\boldsymbol{\mu}}_{k}=[\mu_{1k};\mu_{2k};\ldots;\mu_{Mk}]$
the vector of power control coefficients associated with user $k$.
Thus we can write $\boldsymbol{\nu}_{ik}\trans\boldsymbol{\eta}_{i}$
as $\bar{\boldsymbol{\nu}}_{ik}\boldsymbol{\bar{\mu}}_{k}$, where
\begin{equation}
\bar{\boldsymbol{\nu}}_{ik}\triangleq\Bigl|\boldsymbol{\psi}_{i}\herm\boldsymbol{\psi}_{k}\Bigr|\Bigl[\sqrt{\nu_{1i}}\frac{\beta_{1k}}{\beta_{1i}};\sqrt{\nu_{2i}}\frac{\beta_{2k}}{\beta_{2i}};\ldots;\sqrt{\nu_{Mi}}\frac{\beta_{Mk}}{\beta_{Mi}}\Bigr].
\end{equation}
Similarly, we can write $\mathbf{D}_{ik}\bar{\boldsymbol{\eta}}_{i}$
as $\bar{\mathbf{D}}_{i}\boldsymbol{\bar{\mu}}_{i}$, where $\bar{\mathbf{D}}_{i}$
is the diagonal matrix with the $m$-th diagonal entry equal to $\sqrt{\beta_{mi}}$.
Now the spectral efficiency of the $k$-th user (in nat/s/Hz) can
be expressed as
\begin{equation}
\mathrm{SE}_{k}(\boldsymbol{\mu})=\bigl(1-\frac{T_{p}}{T_{c}}\bigr)\log\bigl(1+\gamma_{k}(\boldsymbol{\mu})\bigr),\label{eq:SEk_mu}
\end{equation}
where $\gamma_{k}(\boldsymbol{\mu})$ is the SINR of the $k$-th user
given by
\begin{equation}
\gamma_{k}(\boldsymbol{\mu})=\frac{\zeta_{d}(\bar{\boldsymbol{\nu}}_{kk}\trans\boldsymbol{\bar{\mu}}_{k})^{2}}{\zeta_{d}\left(\sum_{i\neq k}^{K}(\bar{\boldsymbol{\nu}}_{ik}\trans\boldsymbol{\bar{\mu}}_{i})^{2}+\frac{1}{N}\sum_{i=1}^{K}\bigl\Vert\bar{\mathbf{D}}_{i}\boldsymbol{\bar{\mu}}_{i}\bigr\Vert_{2}^{2}\right)+\frac{1}{N^{2}}}.
\end{equation}
The total spectral efficiency of the system is defined as 
\begin{equation}
\mathrm{SE}(\boldsymbol{\mu})\triangleq\sum\nolimits _{k=1}^{K}\mathrm{SE}_{k}(\boldsymbol{\mu}).\label{eq:SEsum}
\end{equation}
In this paper, we consider a power constraint at each AP which
is given by $||\boldsymbol{\mu}_{m}||^{2}\leq\frac{1}{N},m=1,2,\ldots,M$,
which follows from \eqref{eq:spc:origin}. For the problem formulation
purpose, we define the following set
\begin{equation}
\mathcal{S}=\left\{ \boldsymbol{\mu}|\boldsymbol{\mu}\geq0;\bigl\Vert\boldsymbol{\mu}_{m}\bigr\Vert^{2}\leq\frac{1}{N},m=1,2,\ldots,M\right\}, 
\end{equation}
which is nothing but the feasible set of the utility maximization
problems to be presented. In this paper, we consider the following
four common power control utility optimization problems \cite{Luo:SpectrumManagement:2008},
namely
\begin{itemize}
	\item The problem of average spectral efficiency maximization (SEmax) 
	\begin{equation}
	\boxed{(\mathcal{P}_{1}):\underset{\boldsymbol{\mu}}{\maximize}\ \Bigl\{(1/K)\sum\nolimits _{k=1}^{K}\mathrm{SE}_{k}(\boldsymbol{\mu})\ |\ \boldsymbol{\mu}\in\mathcal{S}\Bigr\}.}\label{eq:SEmaxproblem}
	\end{equation}
	\item The problem of proportional fairness maximization (PFmax)%
	\begin{comment}
	I use Proportional Fairness because it is more common.
	\end{comment}
	{} 
	\begin{equation}
	\hspace{-0.2cm}\boxed{(\mathcal{P}_{2}):\underset{\boldsymbol{\mu}}{\maximize}\ \Bigl\{\sum\nolimits _{k=1}^{K}\log\mathrm{SE}_{k}(\boldsymbol{\mu})\ |\ \boldsymbol{\mu}\in\mathcal{S}\Bigr\}.}\label{eq:GRmaxproblem}
	\end{equation}
	Note that the above problem is equivalent to maximizing $\bigl(\prod_{k=1}^{K}\mathrm{SE}_{k}(\boldsymbol{\mu})\bigr)^{1/K}$
	over the set $\mathcal{S}$. Thus it is also known as the problem
	of geometric-rate maximization.
	\item The problem of the harmonic rate maximization (HRmax) 
	\begin{equation}
	\hspace{-0.3cm}\boxed{(\mathcal{P}_{3}):\underset{\boldsymbol{\mu}}{\maximize}\ \Bigl\{K\Bigl(\sum_{k=1}^{K}\mathrm{SE}_{k}(\boldsymbol{\mu})^{-1}\Bigr)^{-1}\ \Bigr|\ \boldsymbol{\mu}\in\mathcal{S}\Bigr\}.}\label{eq:HRmaxproblem}
	\end{equation}
	\item The problem of maximizing the minimum rate (MRmax) among all users
	(also known as max-min fairness maximization)
	\begin{equation}
	\boxed{(\mathcal{P}_{4}):\underset{\boldsymbol{\mu}}{\maximize}\ \Bigl\{\min_{1\leq k\leq K}\mathrm{SE}_{k}(\boldsymbol{\mu})\ |\ \boldsymbol{\mu}\in\mathcal{S}\Bigr\}.}\label{eq:minrate}
	\end{equation}
\end{itemize}
We note that the above four problems are noncovex and thus difficult
to solve. For such problems, a pragmatic goal is to derive a low complexity
high-performance solution, rather than a globally optimal solution.
To this end, SCA has proved to be very effective and gradually become
a standard mathematical tool \cite{Ngo2017,Ngo2018EE}. The idea of
SCA is to approximate a non-convex program by a series of convex sub-problems.
In all known solutions for the considered problems or related ones,
interior point methods (through the use of off-the-shelf convex solvers)
are invoked to solve these convex problems \cite{Ngo2018EE,Buzzi2017,Interdonato2019},
which do not scale favorably with the problem size. Thus the existing
solutions are unable to characterize the performance limits of cell-free
massive MIMO systems where the number of APs can be in the order of
thousands, even from an off-line design perspective. In this paper, we propose methods that can tackle this scalability problem. In particular,
our proposed methods are based on first order optimization methods
which are presented in the following section.

\section{Proposed Solutions}

In this section, we present solutions to $(\mathcal{P}_{1})$ to $(\mathcal{P}_{4})$,
using the APG methods, a variant
of the proximal gradient method introduced in \cite{Li2015}. In general
the four considered problems can be written in a compact form as \begin{subequations}\label{eq:modifiedproblem}
	\begin{align}
	\underset{\boldsymbol{\mu}}{\maximize} & \quad f(\boldsymbol{\mu})\\
	\st & \quad\boldsymbol{\mu}\in\mathcal{S},
	\end{align}
\end{subequations} where $f(\boldsymbol{\mu})=(1/K)\sum\limits\nolimits _{k=1}^{K}\mathrm{SE}_{k}(\boldsymbol{\mu})$
for problem $(\mathcal{P}_{1})$, $f(\boldsymbol{\mu})=\sum\nolimits _{k=1}^{K}\log\mathrm{SE}_{k}(\boldsymbol{\mu})$
for problem $(\mathcal{P}_{2})$, $f(\boldsymbol{\mu})=K\bigl(\sum_{k=1}^{K}\mathrm{SE}_{k}(\boldsymbol{\mu})^{-1}\bigr)^{-1}$
for $(\mathcal{P}_{3})$, and $f(\boldsymbol{\mu})=\min_{1\leq k\leq K}\mathrm{SE}_{k}(\boldsymbol{\mu})$
for $(\mathcal{P}_{4})$. We remark that the method described in \cite{Li2015}
concerns the following problem
\begin{equation}
\underset{\mathbf{x}\in\mathbb{R}^{n}}{\maximize}\ \{F(\mathbf{x})\equiv f(\mathbf{x})+g(\mathbf{x})\},\label{eq:nonconvex:gen}
\end{equation}
where $f(\mathbf{x})$ is differentiable (but possibly \emph{nonconvex})
and $g(\mathbf{x})$ can be both nonconvex and \emph{nonsmooth}. Further
assumptions on $f(\mathbf{x})$ are listed below:
\begin{itemize}
	\item A1: $f(\mathbf{x})$ is a proper function with Lipschitz continuous
	gradients. A function $f$ is said to have an $L$-Lipschitz continuous
	gradient if there exists some $L>0$ such that 
	\begin{equation}
	||\nabla f(\mathbf{x})-\nabla f(\mathbf{y})||\leq L||\mathbf{x}-\mathbf{y}||,\forall\mathbf{x},\mathbf{y}.\label{eq:gradLipschitz}
	\end{equation}
	\item A2: $f(\mathbf{x})$ is coercive i.e. $f(\mathbf{x})$ is bounded
	from below and $f(\mathbf{x})\rightarrow\infty$ when $\mathbf{x}\rightarrow\infty$.
\end{itemize}
If we let $g(\mathbf{x})$ be the indicator function of $\mathcal{S}$,
defined as
\begin{equation}
\delta_{\mathcal{S}}(\mathbf{x})=\begin{cases}
0 & \mathbf{x}\in\mathcal{S}\\
+\infty & \mathbf{x}\notin\mathcal{S},
\end{cases}
\end{equation}
then \eqref{eq:nonconvex:gen} is actually equivalent to \eqref{eq:modifiedproblem}.
Furthermore, the proximal operator of $g(\mathbf{x})$ becomes the
Euclidean projection onto $\mathcal{S}$. In the following, we customize
the APG methods to solve the considered problems. In essence, the
APG method moves the current point along the gradient of the objective
with a proper step size and then projects the resulting point onto
the feasible set. This step is repeated until some stopping criterion
is met.
\begin{rem}
	One may ask why we have made a change of variables from $\bar{\boldsymbol{\eta}}$
	to $\boldsymbol{\mu}$ in Section \ref{subsec:Problem-Formulation}.
	The question is relevant since the projection onto the feasible set
	expressed in terms of $\bar{\boldsymbol{\eta}}$ can also be done efficiently.
	However, if the objective function of the four considered problems
	is written as a function of $\bar{\boldsymbol{\eta}}$, then there are two
	difficulties arising. Firstly, the expression of the gradient of the
	objective becomes very complicated. Secondly and more importantly,
	the gradient of the objective is not Lipschitz continuous since the
	term $\sqrt{\eta_{mk}}$ would appear in the denominator of the gradient.
	Note that $\eta_{mk}$ can be zero, which can make the gradient unbounded.
\end{rem}

\subsection{Proposed Solution to $(\mathcal{P}_{1})$}

Since $f(\boldsymbol{\mu})$ for $(\mathcal{P}_{1})$ is differentiable,
the proposed algorithm for solving $(\mathcal{P}_{1})$ follows closely
the monotone APG method in \cite{Li2015}, which is outlined in Algorithm
\ref{alg:mAPG}. In Algorithm \ref{alg:mAPG} $\alpha>0$ is called
the step size which should be sufficiently small to guarantee its
convergence and $L_{f}$ is the Lipschitz constant of $\nabla f(\boldsymbol{\mu})$.
The superscripts in Algorithm \ref{alg:mAPG} denote the iteration
count. Also, the notation $P_{\mathcal{S}}(\mathbf{u})$ denotes the
projection onto $\mathcal{S}$, i.e., 
\[
P_{\mathcal{S}}(\mathbf{u})=\arg\min\Bigl\{||\mathbf{x}-\mathbf{u}||\ |\ \mathbf{x}\in\mathcal{S}\Bigr\}.
\]
Note that we adapt the monotone APG method in \cite{Li2015} for minimization
to the context of maximization for our problems. Specifically, from
a given operating point, we move along the direction of the gradient
of $f(\boldsymbol{\mu})$ with the step size $\alpha$, and then project
the resulting point onto the feasible set. We note that $\mathbf{y}^{n}$
in Step \ref{alg:mAPG:extra} is an extrapolated point which is used
for convergence acceleration. However, unlike APG methods for the
convex case, $\mathbf{y}^{n}$ can be a bad extrapolation, and thus
Step \ref{alg:mAPG:update} is required to fix this issue.
\begin{algorithm}[th!]
	\caption{Accelerated projected gradient algorithm for solving $(\mathcal{P}_{1})$-$(\mathcal{P}_{4})$}
	\label{alg:mAPG}
	
	\begin{algorithmic}[1]
		
		\STATE Input: $\boldsymbol{\mu}^{0}>=0,t_{0}=t_{1}=1,\frac{1}{L_{f}}>\alpha>0$
		
		\STATE $\boldsymbol{\mu}^{1}=\mathbf{z}^{1}=\boldsymbol{\mu}^{0}$
		
		\FOR{ $n=1,2,\ldots$}
		
		\STATE $\mathbf{y}^{n}=\boldsymbol{\mu}^{n}+\frac{t_{n-1}}{t_{n}}(\mathbf{z}^{n}-\boldsymbol{\mu}^{n})+\frac{t_{n-1}-1}{t_{n}}(\boldsymbol{\mu}^{n}-\boldsymbol{\mu}^{n-1})$\label{alg:mAPG:extra}
		
		\STATE $\mathbf{z}^{n+1}=P_{\mathcal{S}}(\mathbf{y}^{n}+\alpha\nabla f(\mathbf{y}^{n}))$
		\label{grady}
		
		\STATE $\mathbf{v}^{n+1}=P_{\mathcal{S}}(\boldsymbol{\mu}^{n}+\alpha\nabla f(\boldsymbol{\mu}^{n}))$\label{gradmu}
		
		\STATE $\boldsymbol{\mu}^{n+1}=\begin{cases}
		\mathbf{z}^{n+1} & f(\mathbf{z}^{n+1})\geq f(\mathbf{v}^{n+1})\\
		\mathbf{v}^{n+1} & \textrm{otherwise}
		\end{cases}$\label{alg:mAPG:update}
		
		\STATE $t_{n+1}=0.5\left(\sqrt{4t_{n}^{2}+1}+1\right)$
		
		\ENDFOR
		
		\STATE Output: $\boldsymbol{\mu}^{n}$
		
	\end{algorithmic}
\end{algorithm}
We now give the details for the two main operations of Algorithm
\ref{alg:mAPG}, namely: the projection onto the feasible set $\mathcal{S}$
and the gradient of $f(\boldsymbol{\mu})$.

\subsubsection{Projection onto $\mathcal{S}$}

We show that the projection in Steps \ref{grady} and \ref{gradmu}
in Algorithm \ref{alg:mAPG} can be done \emph{in parallel} and by
\emph{closed-form expressions}. Recall that for given a $\mathbf{x}\in\mathbb{R}^{MK\times1}$,
$P_{\mathcal{S}}(\mathbf{x})$ is the solution to the following problem
%\begin{subequations}
	\begin{equation}
	\underset{\boldsymbol{\mu}\in\mathbb{R}^{MK\times1}}{\minimize}\ \Bigl\{||\boldsymbol{\mu}-\mathbf{x}||^{2}\ \Bigl|\ \boldsymbol{\mu}\geq0;||\boldsymbol{\mu}_{m}||^{2}\leq\frac{1}{N},	m=1,2,\ldots,M\Bigr\}.
	\end{equation}
%\end{subequations} 
It is easy to see that the above problem can
be decomposed into sub-problems at each AP $m$ as 
	\begin{equation}\label{eq:project:subprob}
	\underset{\boldsymbol{\mu}_{m}\in\mathbb{R}^{K\times1}}{\minimize}  \Bigl\{||\boldsymbol{\mu}_{m}-\mathbf{x}_{m}||^{2}\ \Bigl|\ \boldsymbol{\mu}_{m}\geq0;||\boldsymbol{\mu}_{m}||^{2}\leq\frac{1}{N}\Bigr\}.
	\end{equation}
The above problem is in fact the projection onto
the intersection of a ball and the positive orthant. Interestingly,
the analytical solution to this problem can be found by applying \cite[Theorem 7.1]{Bauschke2017},
which produces
\begin{equation}
\boldsymbol{\mu}_{m}=\frac{\sqrt{1/N}}{\max\left(\sqrt{1/N},||[\mathbf{x}_{m}]_{+}||\right)}[\mathbf{x}_{m}]_{+}.\label{eq:projectionEuclidean}
\end{equation}
The above expression means that we first project $\mathbf{x}_{m}$
onto the positive orthant and then onto the Euclidean ball of radius
$\sqrt{1/N}$. A simpler way to prove \eqref{eq:projectionEuclidean}
is detailed in Appendix \ref{apx:projection}.

\subsubsection{Gradient of $f(\boldsymbol{\mu})$ for $(\mathcal{P}_{1})$}

To implement Algorithm \ref{alg:mAPG}, we also need to compute $\nabla_{\boldsymbol{\mu}}f(\boldsymbol{\mu})$,
which is derived in what follows. We know that the gradient of a multi-variable
function is the vector of all its partial derivatives, i.e.
\begin{align}
\nabla f(\boldsymbol{\mu}) & =\left[\frac{\partial}{\partial\bar{\boldsymbol{\mu}}_{1}}f(\boldsymbol{\mu});\frac{\partial}{\partial\bar{\boldsymbol{\mu}}_{2}}f(\boldsymbol{\mu}),\ldots,\frac{\partial}{\partial\bar{\boldsymbol{\mu}}_{K}}f(\boldsymbol{\mu})\right],\label{eq:gradfmu}
\end{align}
where $\frac{\partial}{\partial\bar{\boldsymbol{\mu}}_{i}}f(\boldsymbol{\mu})=(1/K)\sum_{k=1}^{K}\frac{\partial}{\partial\bar{\boldsymbol{\mu}}_{i}}\mathrm{SE}_{k}(\boldsymbol{\mu})$.
Thus, it basically boils down to finding $\frac{\partial}{\partial\bar{\boldsymbol{\mu}}_{i}}\mathrm{SE}_{k}(\boldsymbol{\mu})$.
Let us define $b_{k}(\boldsymbol{\mu})=\zeta_{d}(\bar{\boldsymbol{\nu}}_{kk}\trans\boldsymbol{\bar{\mu}}_{k})^{2}$
and $c_{k}(\boldsymbol{\mu})=\zeta_{d}\left(\sum_{i\neq k}^{K}(\bar{\boldsymbol{\nu}}_{ik}\trans\boldsymbol{\bar{\mu}}_{i})^{2}+\frac{1}{N}\sum_{i=1}^{K}||\bar{\mathbf{D}}_{i}\boldsymbol{\bar{\mu}}_{i}||_{2}^{2}\right)+\frac{1}{N^{2}}$.
Then we can rewrite $\mathrm{SE}_{k}(\boldsymbol{\mu})$ as
\begin{equation}
\mathrm{SE}_{k}(\boldsymbol{\mu})=\log\bigl(b_{k}(\boldsymbol{\mu})+c_{k}(\boldsymbol{\mu})\bigr)-\log c_{k}(\boldsymbol{\mu}).\label{eq:SElogterm}
\end{equation}
The gradient of $\mathrm{SE}_{k}(\boldsymbol{\mu})$ with respect
to $\boldsymbol{\bar{\mu}}_{i}$, $i=1,2,\ldots,K$, is found as 
\begin{equation}
\frac{\partial}{\partial\bar{\boldsymbol{\mu}}_{i}}\mathrm{SE}_{k}(\boldsymbol{\mu})=\frac{\frac{\partial}{\partial\bar{\boldsymbol{\mu}}_{i}}\left(b_{k}(\boldsymbol{\mu})+c_{k}(\boldsymbol{\mu})\right)}{b_{k}(\boldsymbol{\mu})+c_{k}(\boldsymbol{\mu})}-\frac{\frac{\partial}{\partial\bar{\boldsymbol{\mu}}_{i}}c_{k}(\boldsymbol{\mu})}{c_{k}(\boldsymbol{\mu})}\label{eq:gradGamma}
\end{equation}
Now we recall the following identity $\nabla||\mathbf{A}\mathbf{x}||^{2}=2\mathbf{A}\trans\mathbf{A}\mathbf{x}$
for any symmetric matrix $\mathbf{A}$, and thus $\nabla_{\boldsymbol{\bar{\mu}}_{i}}b_{k}(\boldsymbol{\mu})$
and $\nabla_{\boldsymbol{\bar{\mu}}_{i}}c_{k}(\boldsymbol{\mu})$
are respectively given by 
\begin{equation}
\frac{\partial}{\partial\bar{\boldsymbol{\mu}}_{i}}b_{k}(\boldsymbol{\mu})=\begin{cases}
2\zeta_{d}\bar{\boldsymbol{\nu}}_{kk}\bar{\boldsymbol{\nu}}_{kk}\trans\boldsymbol{\bar{\mu}}_{k}, & i=k\\
0, & i\neq k
\end{cases}\label{eq:BkGradient}
\end{equation}

\begin{equation}
\frac{\partial}{\partial\bar{\boldsymbol{\mu}}_{i}}c_{k}(\boldsymbol{\mu})=\begin{cases}
2(\zeta_{d}/N)\mathbf{\bar{D}}_{k}^{2}\boldsymbol{\bar{\mu}}_{k}, & i=k\\
2\zeta_{d}\bar{\boldsymbol{\nu}}_{ik}\bar{\boldsymbol{\nu}}_{ik}\trans\boldsymbol{\bar{\mu}}_{i}+\frac{2\zeta_{d}}{N}\mathbf{\bar{D}}_{i}^{2}\boldsymbol{\bar{\mu}}_{i}, & i\neq k
\end{cases}.\label{eq:CkGradient}
\end{equation}

\subsection{Improved Convergence with Line Search}

For $(\mathcal{P}_{1})$, from \eqref{eq:gradGamma}, \eqref{eq:BkGradient},
and \eqref{eq:CkGradient}, it easy to check that $\nabla f(\boldsymbol{\mu})$
is Lipschitz continuous, or equivalently $f(\boldsymbol{\mu})$ has
Lipschitz continuous gradient. That is, there exists a constant $L_{f}>0$
such that 
\begin{equation}
||\nabla f(\mathbf{x})-\nabla f(\mathbf{\mathbf{y}})||\leq L_{f}||\mathbf{x}-\mathbf{y}||\ \forall\mathbf{x},\mathbf{y}\in\mathcal{S}.\label{eq:gradLips-1}
\end{equation}
Further details are given in Appendix \ref{apx:LipschitzConstant}.

In practice, we in fact do not need to compute a Lipschitz constant
of $\nabla f(\boldsymbol{\mu})$ for two reasons. First, the best
Lipschitz constant of $\nabla f(\boldsymbol{\mu})$ (i.e. the smallest
$L$ such that \eqref{eq:gradLips-1} holds) is hard to find. Second,
the conditions $\alpha<\frac{1}{L_{f}}$ is sufficient but not necessary
for Algorithm \ref{alg:mAPG} to converge. Thus, we can allow $\alpha$
to take on larger values to speed up the convergence of Algorithm
\ref{alg:mAPG} by means of a linear search procedure.%
\begin{comment}
We do not use the Armijo line search in our paper.
\end{comment}
{} In this paper we use line search with the Barzilai-Borwein (BB) rule
to compute step sizes for Algorithm \ref{alg:mAPG}. The APG method
with line search backtracking line search is summarized in Algorithm
\ref{alg:mAPG_LS}. The step sizes $\alpha_{y}$ and $\alpha_{\mu}$
computed in Steps \ref{alg:mAPG_LS:stepsizey} and \ref{alg:mAPG_LS:stepsizemu}
can be viewed as local estimate of the optimal Lipschitz constant
of the gradient at $\mathbf{y}^{n-1}$ and $\boldsymbol{\mu}^{n-1}$,
respectively.
\begin{algorithm}[th!]
	\caption{APG method with line search for solving $(\mathcal{P}_{1})$-$(\mathcal{P}_{4})$}
	\label{alg:mAPG_LS}
	
	\begin{algorithmic}[1]
		
		\STATE Input: $\boldsymbol{\mu}^{0}>=0,t_{0}=0,t_{1}=1,\alpha_{\mu}>0,\alpha_{y}>0,\delta>0,\rho<1$
		
		\STATE $\boldsymbol{\mu}^{1}=\mathbf{z}^{1}=\boldsymbol{\mu}^{0}$
		
		\FOR{ $n=1,2,\ldots$}
		
		\STATE $\mathbf{y}^{n}=\boldsymbol{\mu}^{n}+\frac{t_{n-1}}{t_{n}}(\mathbf{z}^{n}-\boldsymbol{\mu}^{n})+\frac{t_{n-1}-1}{t_{n}}(\boldsymbol{\mu}^{n}-\boldsymbol{\mu}^{n-1})$
		
		\STATE $\mathbf{s}^{n}=\mathbf{z}^{n}-\mathbf{y}^{n-1},\mathbf{r}^{n}=\nabla f(\mathbf{z}^{n})-\nabla f(\mathbf{y}^{n-1})$
		
		\STATE $\alpha_{y}=\left\langle \mathbf{s}^{n},\mathbf{s}^{n}\right\rangle /\left\langle \mathbf{s}^{n},\mathbf{r}^{n}\right\rangle \ \mathrm{or}\ \alpha_{y}=\left\langle \mathbf{s}^{n},\mathbf{r}^{n}\right\rangle /\left\langle \mathbf{r}^{n},\mathbf{r}^{n}\right\rangle $\label{alg:mAPG_LS:stepsizey}
		
		\STATE $\mathbf{s}^{n}=\mathbf{v}^{n}-\boldsymbol{\mu}^{n-1},\mathbf{r}^{n}=\nabla f(\mathbf{v}^{n})-\nabla f(\boldsymbol{\mu}^{n-1})$
		
		\STATE $\alpha_{\mu}=\left\langle \mathbf{s}^{n},\mathbf{s}^{n}\right\rangle /\left\langle \mathbf{s}^{n},\mathbf{r}^{n}\right\rangle \ \mathrm{or}\ \alpha_{\mu}=\left\langle \mathbf{s}^{n},\mathbf{r}^{n}\right\rangle /\left\langle \mathbf{r}^{n},\mathbf{r}^{n}\right\rangle $\label{alg:mAPG_LS:stepsizemu}
		
		\REPEAT
		
		\STATE $\mathbf{z}^{n+1}=P_{\mathcal{S}}(\mathbf{y}^{n}+\alpha_{y}\nabla f(\mathbf{y}^{n}))$
		
		\STATE $\alpha_{y}=\alpha_{y}\times\rho$
		
		\UNTIL $F(\mathbf{z}^{n+1})\geq F(\mathbf{y}^{n})+\delta||\mathbf{z}^{n+1}-\mathbf{y}^{n}||^{2}$
		
		\REPEAT
		
		\STATE $\mathbf{v}^{n+1}=P_{\mathcal{S}}(\boldsymbol{\mu}^{n}+\alpha_{\mu}\nabla f(\boldsymbol{\mu}^{n}))$
		
		\STATE $\alpha_{\mu}=\alpha_{\mu}\times\rho$
		
		\UNTIL $F(\mathbf{v}^{n+1})\geq F(\boldsymbol{\mu}^{n})+\delta||\mathbf{v}^{n+1}-\boldsymbol{\mu}^{n}||^{2}$
		
		\STATE $\boldsymbol{\mu}^{n+1}=\begin{cases}
		\mathbf{z}^{n+1} & F(\mathbf{z}^{n+1})\geq F(\mathbf{v}^{n+1})\\
		\mathbf{v}^{n+1} & \textrm{otherwise}
		\end{cases}$
		
		\STATE $t_{n+1}=0.5\left(\sqrt{4t_{n}^{2}+1}+1\right)$
		
		\ENDFOR
		
		\STATE Output: $\boldsymbol{\mu}^{n}$
		
	\end{algorithmic}
\end{algorithm}

\subsection{Customization for Solving $(\mathcal{P}_{2})$ and $(\mathcal{P}_{3})$}

We remark that Algorithms \ref{alg:mAPG} and \ref{alg:mAPG_LS} are
unified in the sense that they can be easily modified to solve the
remaining considered problems. In this subsection we explain how to
apply Algorithms \ref{alg:mAPG} and \ref{alg:mAPG_LS} to solve problems
$(\mathcal{P}_{2})$ and $(\mathcal{P}_{3})$. First note that the
objective functions in $(\mathcal{P}_{2})$ and $(\mathcal{P}_{3})$
are differentiable and the application of Algorithms \ref{alg:mAPG}
and \ref{alg:mAPG_LS} is straightforward. Specifically, for the
PFmax problem (i.e. $(\mathcal{P}_{2})$), the objective is
\begin{align}
f(\boldsymbol{\mu}) & =\sum\nolimits _{k=1}^{K}\log\mathrm{SE}_{k}(\boldsymbol{\mu}),
\end{align}
and thus $\nabla f(\boldsymbol{\mu})$ is given by \eqref{eq:gradfmu},
where $\nabla_{\bar{\boldsymbol{\mu}}_{i}}f(\boldsymbol{\mu})$ is
found as
\begin{equation}
\frac{\partial}{\partial\bar{\boldsymbol{\mu}}_{i}}f(\boldsymbol{\mu})=\sum_{k=1}^{K}\frac{1}{\mathrm{SE}_{k}(\boldsymbol{\mu})}\frac{\partial}{\partial\bar{\boldsymbol{\mu}}_{i}}\mathrm{SE}_{k}(\boldsymbol{\mu}),\label{eq:gradP2}
\end{equation}
where $\frac{\partial}{\partial\bar{\boldsymbol{\mu}}_{i}}\mathrm{SE}_{k}(\boldsymbol{\mu})$
is provided in \eqref{eq:gradGamma}. For the HRmax problem (i.e.
$(\mathcal{P}_{2})$), the objective is%
\begin{comment}
Note that we now consider maximization, so we don't need to negate
the objective.
\end{comment}
\begin{equation}
f(\boldsymbol{\mu})=K\Bigl(\sum_{k=1}^{K}\mathrm{SE}_{k}(\boldsymbol{\mu})^{-1}\Bigr)^{-1}.
\end{equation}
The gradient of the function can be found similarly where $\nabla_{\bar{\boldsymbol{\mu}}_{i}}f(\boldsymbol{\mu})$
is written as
\begin{equation}
\frac{\partial}{\partial\bar{\boldsymbol{\mu}}_{i}}f(\boldsymbol{\mu})=K\Bigl(\sum_{k=1}^{K}\mathrm{SE}_{k}(\boldsymbol{\mu})^{-1}\Bigr)^{-2}\sum_{k=1}^{K}\frac{1}{\bigl(\mathrm{SE}_{k}(\boldsymbol{\mu})\bigr)^{2}}\frac{\partial}{\partial\bar{\boldsymbol{\mu}}_{i}}\mathrm{SE}_{k}(\boldsymbol{\mu}).\label{eq:gradP3}
\end{equation}
where $\frac{\partial}{\partial\bar{\boldsymbol{\mu}}_{i}}\mathrm{SE}_{k}(\boldsymbol{\mu})$
is again provided in \eqref{eq:gradGamma}.
\begin{rem}
	For $(\mathcal{P}_{2})$ and $(\mathcal{P}_{3})$, the utility functions
	are not Lipschitz continuous gradient in principle since the data rate
	$\mathrm{SE}_{k}(\boldsymbol{\mu})$ can be zero for   some user $k$.
	Consequently, the gradient of the objective becomes unbounded due
	to the term $\frac{1}{\mathrm{SE}_{k}(\boldsymbol{\mu})}$ in \eqref{eq:gradP2}
	and \eqref{eq:gradP3}. In practice, to fix this problem we simply
	add a fixed regularization parameter $\epsilon$ (say, $\epsilon=10^{-6}$)
	and consider $f(\boldsymbol{\mu})=\sum\nolimits _{k=1}^{K}\log\bigl(\epsilon+\mathrm{SE}_{k}(\boldsymbol{\mu})\bigr)$
	for $(\mathcal{P}_{2})$ and $f(\boldsymbol{\mu})=K\Bigl(\sum_{k=1}^{K}\bigl(\epsilon+\mathrm{SE}_{k}(\boldsymbol{\mu})\bigr)^{-1}\Bigr)^{-1}$
	for $(\mathcal{P}_{3})$. In this way $f(\boldsymbol{\mu})$ is Lipschitz
	continuous gradient.
\end{rem}

\subsection{Proposed Solution to $(\mathcal{P}_{4})$}

Problem $(\mathcal{P}_{4})$ deserves further discussions since the
objective is nonsmooth. We recall that for $(\mathcal{P}_{4})$ the
objective function is
\begin{equation}
f(\boldsymbol{\mu})=\underset{1\leq k\leq K}{\min}\mathrm{SE}_{k}(\boldsymbol{\mu}),
\end{equation}
which is \emph{non-differentiable}. Thus a straightforward application
of the APG method is impossible. To overcome this issue, we adopt a
smoothing technique. In particular, $f(\boldsymbol{\mu})$ is approximated
by the following log-sum-exp function given by\cite{Nesterov2005a}
\begin{equation}
f_{\tau}(\boldsymbol{\mu})=-\frac{1}{\tau}\log\bigl(\frac{1}{K}\sum\nolimits _{k=1}^{K}\exp\bigl(-\tau\mathrm{SE}_{k}(\boldsymbol{\mu})\bigr),\label{eq:smoothapprox}
\end{equation}
where $\tau>0$ is the positive \emph{smoothness} parameter. To obtain
\eqref{eq:smoothapprox}, we have used the fact that $\underset{1\leq k\leq K}{\min}\mathrm{SE}_{k}(\boldsymbol{\mu})=-\underset{1\leq k\leq K}{\max}(-\mathrm{SE}_{k}(\boldsymbol{\mu}))$.
It was proved in \cite{Nesterov2005a} that $f(\boldsymbol{\mu})+\frac{\log K}{\tau}\geq f_{\tau}(\boldsymbol{\mu})\geq f(\boldsymbol{\mu})$.
In other words, $f_{\tau}(\boldsymbol{\mu})$ is a differentiable
approximation of $f(\boldsymbol{\mu})$ with a numerical accuracy
of $\frac{\log K}{\tau}$. Thus, with a sufficiently high $\tau$,
we can find an approximate solution to $(\mathcal{P}_{4})$ by running
Algorithm \ref{alg:mAPG} with $f(\boldsymbol{\mu})$ being replaced
by $f_{\tau}(\boldsymbol{\mu})$ in \eqref{eq:smoothapprox}. In this
regard, the gradient of $f_{\tau}(\boldsymbol{\mu})$ is easily found
as $\nabla_{\bar{\boldsymbol{\mu}}}f_{\tau}(\boldsymbol{\mu})=[\frac{\partial}{\partial\bar{\boldsymbol{\mu}}_{1}}f_{\tau}(\boldsymbol{\mu}),\frac{\partial}{\partial\bar{\boldsymbol{\mu}}_{2}}f_{\tau}(\boldsymbol{\mu}),\ldots,\frac{\partial}{\partial\bar{\boldsymbol{\mu}}_{K}}f_{\tau}(\boldsymbol{\mu})]$
where $\frac{\partial}{\partial\bar{\boldsymbol{\mu}}_{i}}f_{\tau}(\boldsymbol{\mu})$
is given by 
\begin{equation}
\frac{\partial}{\partial\bar{\boldsymbol{\mu}}_{i}}f_{\tau}(\boldsymbol{\mu})=\frac{\sum_{k=1}^{K}\Bigl(\exp\bigl(-\tau\mathrm{SE}_{k}(\boldsymbol{\mu})\bigr)\frac{\partial}{\partial\bar{\boldsymbol{\mu}}_{k}}\mathrm{SE}_{k}(\boldsymbol{\mu})\Bigr)}{\sum_{k=1}^{K}\exp\bigl(-\tau\mathrm{SE}_{k}(\boldsymbol{\mu})\bigr)}.
\end{equation}

\section{Complexity and Convergence Analysis of Proposed Methods}

\subsection{Complexity Analysis\label{subsec:Complexity-Analysis}}

We now provide the complexity analysis of the proposed algorithms
for one iteration using the big-O notation. It is clear that the complexity
of Algorithm~\ref{alg:mAPG} is dominated by the computation of
three quantities: the objective, the gradient, and the projection.
It is easy to see that $KM$ multiplications are required to compute
$\mathrm{SE}_{k}(\boldsymbol{\mu})$. Therefore, the complexity of
finding $f(\boldsymbol{\mu})=\sum_{k=1}^{K}\mathrm{SE}_{k}(\boldsymbol{\mu})$
is $\mathcal{O}(K^{2}M).$ In a similar way, we can find that the
complexity of $\nabla f(\boldsymbol{\mu})$ is also $\mathcal{O}(K^{2}M)$.
The projection of $\boldsymbol{\mu}$ onto $\mathcal{S}$ is given
in \eqref{eq:projectionEuclidean}, which requires the computation
of the $l_{2}$-norm of $K\times1$ vector $\mathbf{x}_{m}$ at each
AP, and thus the complexity of the projection is $\mathcal{O}(KM)$.
In summary, the per-iteration complexity of the proposed algorithm
for solving $(\mathcal{P}_{1})$ is $\mathcal{O}(K^{2}M)$. Similarly,
the per-iteration complexity for solving $(\mathcal{P}_{2})$, $(\mathcal{P}_{3})$ and $(\mathcal{P}_{4})$
is also $\mathcal{O}(K^{2}M)$.

To appreciate the low-complexity of the proposed methods, we now provide
the per-iteration complexity of the SCA-based method for solving $(\mathcal{P}_{1})$
derived from \cite{Ngo2018EE}. Note that the iterative method presented
in \cite{Ngo2018EE} is dedicated to the problem of total energy efficiency
maximization but it can be easily customized to solve $(\mathcal{P}_{1})$.
Specifically, the convex sub-problem at the iteration $n+1$ of the
SCA-based method reads\begin{subequations}\label{eq:SCA:subprob}
	\begin{align}
	\underset{\boldsymbol{\mu}\geq0,\mathbf{t}\geq0}{\minimize} & \ \bigl(t_{1}t_{2}\ldots t_{K}\bigr)^{1/K}\\
	\st & \ F(\boldsymbol{\mu},t_{k};\boldsymbol{\mu}^{n},t_{k}^{n})\geq\zeta_{d}N^{2}\sum_{i\neq k}^{K}(\bar{\boldsymbol{\nu}}_{ik}\trans\boldsymbol{\bar{\mu}}_{i})^{2} +\zeta_{d}N\sum_{k=1}^{K}\bigl\Vert\bar{\mathbf{D}}_{k}\boldsymbol{\bar{\mu}}_{k}\bigr\Vert_{2}^{2}+1,\ k=1,\ldots,K\\
	& \ \left\Vert \boldsymbol{\mu}_{m}\right\Vert ^{2}\leq\frac{1}{N},\ m=1,\ldots,M,
	\end{align}
\end{subequations}where
\begin{equation}
F(\boldsymbol{\mu},t_{k};\boldsymbol{\mu}^{n},t_{k}^{n})=f(\boldsymbol{\mu}^{n},t_{k}^{n})+
\nabla_{\boldsymbol{\mu}}f(\boldsymbol{\mu}^{n},t_{k}^{n})\trans(\boldsymbol{\mu}-\boldsymbol{\mu}^{n})+\partial_{t_{k}}f(\boldsymbol{\mu}^{n},t_{k}^{n})(t_{k}-t_{k}^{n}),
\end{equation}
and 
\[
f(\boldsymbol{\mu},t_{k})\triangleq\frac{\zeta_{d}N^{2}\sum_{i=1}^{K}(\bar{\boldsymbol{\nu}}_{ik}\trans\boldsymbol{\bar{\mu}}_{i})^{2}+\zeta_{d}N\sum_{k=1}^{K}\bigl\Vert\bar{\mathbf{D}}_{k}\boldsymbol{\bar{\mu}}_{k}\bigr\Vert_{2}^{2}+1}{t_{k}}.
\]
We remark that the objective admits a second order cone reformulation
and thus \eqref{eq:SCA:subprob} is a second order cone program. According
to \cite[Sect. 6.6.2]{BenTal:2011}, the complexity to solve \eqref{eq:SCA:subprob}
is $\mathcal{O}\bigl(\sqrt{K+M}M^{3}K^{4}\bigr)$, which is much larger
than $\mathcal{O}(K^{2}M)$ for the proposed method, especially when $M$ and $K$ are large.

\subsection{Convergence Analysis}

We now discuss the convergence result of Algorithms \ref{alg:mAPG}
and \ref{alg:mAPG_LS} for solving $(\mathcal{P}_{1})$, which is
stated in the following lemma.
\begin{lem}
	Let $\{\boldsymbol{\mu}^{n}\}$ be the sequence produced by Algorithms
	\ref{alg:mAPG} or \ref{alg:mAPG_LS}. Then the following properties
	hold.
	\begin{itemize}
		\item The sequence of the objective values $\{f(\boldsymbol{\mu}^{n})\}$
		is nondecreasing and convergent.
		\item The sequence $\{\boldsymbol{\mu}^{n}\}$ is bounded and any limit
		point of $\{\boldsymbol{\mu}^{n}\}$ is a critical point of $(\mathcal{P}_{1})$.
	\end{itemize}
\end{lem}
\begin{IEEEproof}
	Please see Appendix \ref{sec:Convergence-Proof}.
\end{IEEEproof}
The same convergence result applies to Algorithms \ref{alg:mAPG}
and \ref{alg:mAPG_LS} for solving $(\mathcal{P}_{2})$, $(\mathcal{P}_{3})$
and $(\mathcal{P}_{4})$.

\section{\label{sec:Numerical-Results}Numerical Results}

In this section, we evaluate the performance of the proposed methods
in terms of computational complexity and achieved spectral efficiency.
All simulations results are obtained using Algorithm \ref{alg:mAPG_LS}
since it has a faster convergence rate. The users and the APs are
uniformly dropped over a $D\times D$ \si{\km\squared}. The large-scale
fading coefficient between the $m$-th AP and the $k$-th user is
generated as 
\[
\beta_{mk}=\mathrm{PL}_{mk}.z_{mk},
\]
where $\mathrm{PL}_{mk}$ and $z_{mk}$ represent the path loss and
log-normal shadowing with mean zero and standard deviation $\sigma_{\textrm{sh}}$,
respectively. In this paper, we adopt the three-slope path loss model
as in \cite{Ngo2018EE}, in which $\mathrm{PL}_{mk}$ (in dB) equals  $-L-15\log_{10}(d_{1})-20\log_{10}(d_{0})$ if $d_{mk}<d_{0}$, equals $-L-15\log_{10}(d_{1})-20\log_{10}(d_{mk})$ if  $d_{0}<d_{mk}<d_{1}$, and equals $-L-35\log_{10}(d_{mk})$ otherwise,
where $L$ is a constant dependent on carrier frequency, $d_{mk}$
(in \si{\km}) is the distance between the $m$-th AP and the $k$-th
user, and $d_{0}$ and $d_{1}$ (both in \si{\km}) are reference
distances. Similar to \cite{Ngo2018EE}, we choose $L=140.7$ dB,
$d_{0}=$ \SI{0.01}{\km} and $d_{1}=$ \SI{0.05}{\km}.
We consider a system having a bandwidth of $B=$ \SI{20}{\MHz},
the noise power density is $N_{0}=\text{\textminus}174$ (dBm/\si{\hertz}),
and a noise figure of $9$ dB. If not otherwise mentioned, the length of the coherence interval and
the uplink training phase are set to $T_{p}=20$, $T_{c}=200$, respectively.
 We set the power transmit power for downlink
data transmission and uplink training phase (before normalization)
as $\zeta_{d}=1$ W and $\zeta_{p}=0.2$ W, respectively.%
\begin{comment}
We do consider multi-antenna APS
\end{comment}
{} These parameters are taken from \cite{Ngo2018EE}.

In the first numerical experiment, we compare the convergence rate
of the proposed method with the SCA-based method presented in \cite{Ngo2018EE}
as explained in Section \ref{subsec:Complexity-Analysis}. To solve
\eqref{eq:SCA:subprob}, we use convex conic solver MOSEK \cite{MOSEKApS15}
through the modeling tool YALMIP \cite{YALMIP}.
\begin{figure}[tbh]
	\begin{centering}
	\subfloat[SEmax (i.e. $(\mathcal{P}_{1})$)]{\label{fig:convergence-1}\includegraphics[bb=62bp 559bp 294bp 741bp]{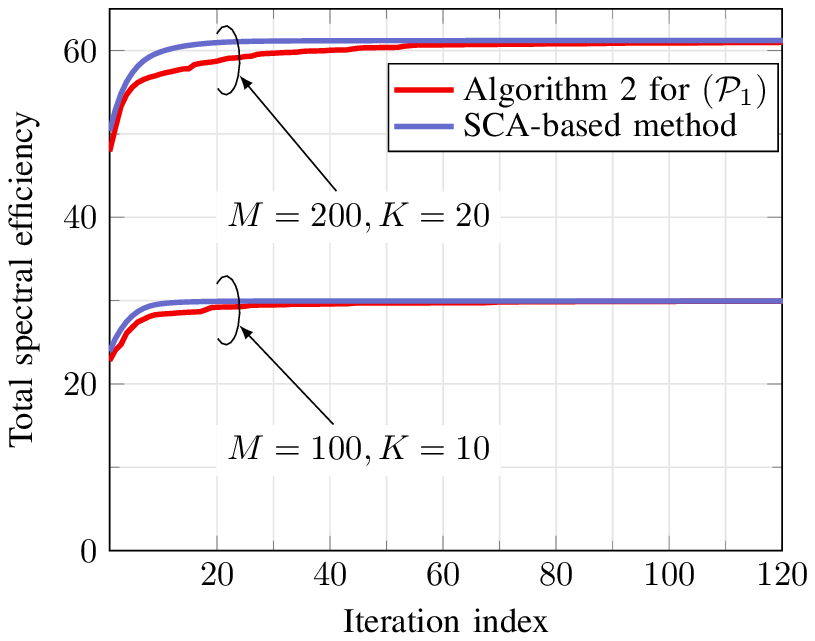}}
	\textsf{}\subfloat[Max-min fairness (i.e. $(\mathcal{P}_{4})$)]{\label{fig:convergence-2}\includegraphics[bb=62bp 557bp 300bp 741bp]{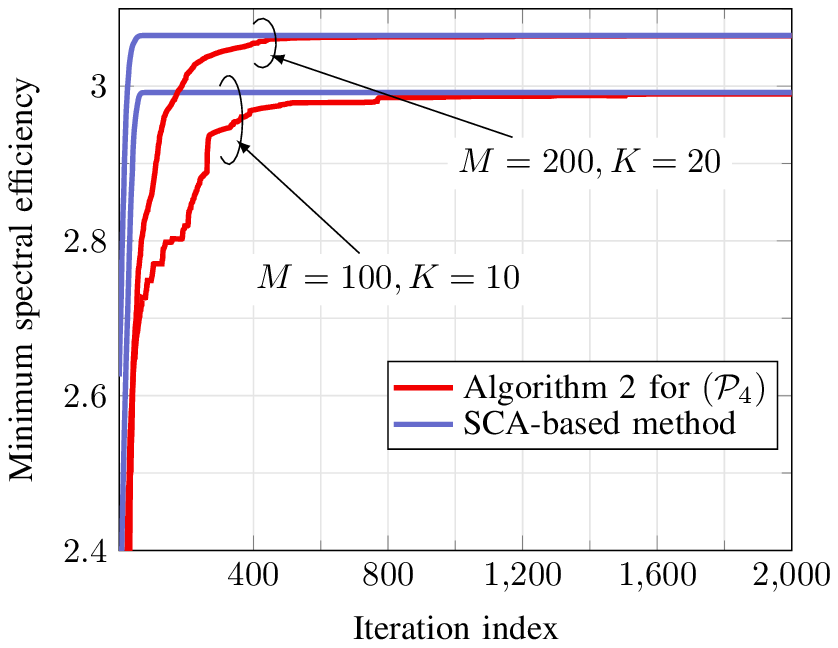}}
		
	\caption{Total SE and the minimum SE versus the number of iterations. The values of $M$ and $K$ are given explicitly the figure. Each AP is equipped with a single antenna.}
	\label{fig:convergence}
	\end{centering}
\end{figure}
In particular, Figs.~\ref{fig:convergence-1} and \ref{fig:convergence-2}
show the convergence of Algorithm \ref{alg:mAPG_LS} and the SCA-based
method for the total spectral efficiency and the min-rate maximization
problem, respectively. We can see that Algorithm \ref{alg:mAPG_LS}
and SCA-based methods achieve the same performance but the SCA-based
method requires fewer iterations to return a solution. However, the
main advantage of Algorithm \ref{alg:mAPG_LS} over the SCA-based
method is that each iteration of the proposed method is very memory
efficient and can be done by closed-form expressions, and hence, is
executed very fast. As a result, the total run-time of the proposed
method is far less than that of the SCA-based method as shown in Table~\ref{table: Table 1}.
In Table~\ref{table: Table 1}, we report the actual run-time of
both methods to solve the SEmax problem. Here, we execute our codes
on a 64-bit Windows operating system with 16 GB RAM and Intel CORE
i7, 3.7 GHz. Both iterative methods are terminated when the difference
of the objective for the last $5$ iterations is less than $10^{-3}$.
\begin{table}[tbh]
	\caption{Comparison of run-time (in seconds) between Algorithm \ref{alg:mAPG_LS}
		and the SCA-based method for solving the SEmax problem. The values
		of other parameters are taken as $K=40,N=1$ and $D=1$.}
	\label{table: Table 1}
	\centering{}%
	\begin{tabular}{c|c|c}
		\hline 
		number of APs & SCA-based method & Algorithm \ref{alg:mAPG_LS}\tabularnewline
		\hline 
		200 & 330.84 & \textbf{2.88}\tabularnewline
		\hline 
		400 & 408.94 & \textbf{9.42}\tabularnewline
		\hline 
		800 & 1115.18 & \textbf{18.07}\tabularnewline
		\hline 
		1600 & 1648.09 & \textbf{49.45}\tabularnewline
		\hline 
	\end{tabular}
\end{table}

\begin{figure}[tbh]
	\begin{centering}
		\includegraphics[bb=63bp 556bp 303bp 741bp]{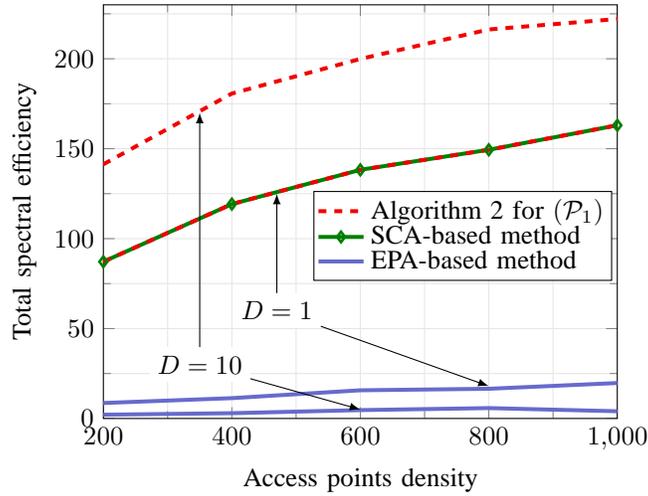}
		\par\end{centering}
	\caption{Total spectral efficiency versus AP density. The number of users is
		$K=40$.}
	\label{fig:SE:APdensity}
\end{figure}

We next take advantage of the proposed methods to explore the spectral
efficiency performance of large-scale cell-free massive MIMO that
can cover, e.g. a large metropolitan area in our vision. In particular,
we investigate the performance of cell-free massive MIMO for two cases
$D=1$ and $D=10$. To obtain a fair comparison, we keep the AP density (defined as the number of APs per square kilometer) same for both
cases. Note that the AP density of \num{1000} means there are \num{10000}
APs for the case of $D=10$, which has not been studied in the literature
previously. To appreciate the proposed method for this large-scale
scenario, we compare it with the SCA method and the equal power allocation
(EPA) method where the power control coefficient $\eta_{mk}$ is given
by, $\eta_{mk}=(\sum_{i=1}^{k}\nu_{mi})^{-1}$. The results in Fig.~\ref{fig:SE:APdensity}
are interesting. First, increasing the AP density expectedly improves
the total spectral efficiency of the system. Second, for the same
AP density, a larger area provides a better sum spectral efficiency.
The reason is that for a larger area, the users that are served by
the APs become far apart each other. As a result, the inter-user inference
becomes weaker, leading to an improved sum spectral efficiency. On
the other hand, the EPA method yields smaller spectral efficiency
as the coverage area is larger because more power should be to spent
to the users having small path loss. The SCA-based method produces
the same spectral efficiency as the proposed APG method but it is
unable to run for $D=10$ on the system specifications mentioned above.
Thus, the proposed scheme outperforms both SCA-based and the EPA methods
in terms of sum spectral efficiency and coverage area. 

\begin{figure}[tbh]
	\begin{centering}
		\textsf{\includegraphics[bb=62bp 552bp 303bp 738bp]{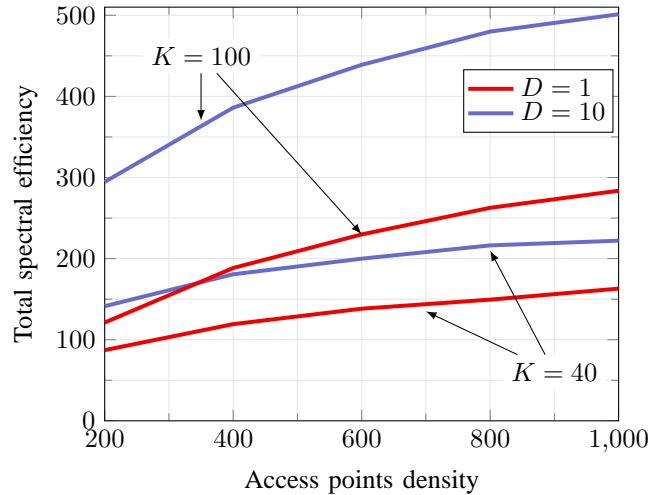}}
		\par\end{centering}
	\caption{Total spectral efficiency versus AP density and number of users.}
	\label{fig:SEvsMK}
\end{figure}
In Fig. \ref{fig:SEvsMK}, we again investigate the spectral efficiency
performance of the AP density but for different number of users (i.e.
$K=100$ and $K=40$). It can be seen that for a given AP density,
the SE is increased if the number of users becomes larger. The gain
is more profound for larger AP density due to the fact that more APs
allow for more efficient exploitation of multiuser diversity gains.

\begin{figure}[tbh]
	\begin{centering}
		
	\subfloat[Small-scale problem: $K=20$ and $M=100$]{\includegraphics[bb=62bp 551bp 293bp 738bp]{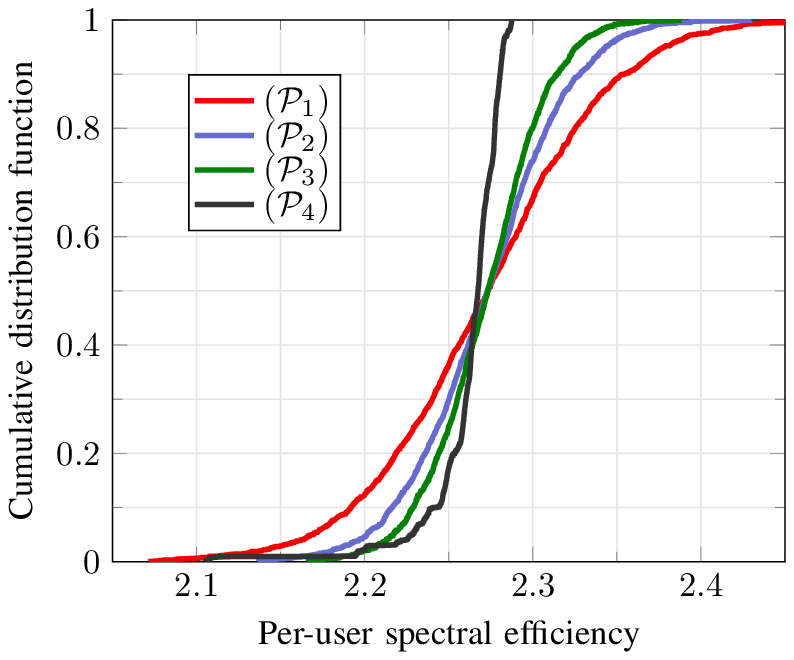}\label{fig:CDFK20}}
	\subfloat[Large-scale problem: $K=500$ and $M=2000$]{\includegraphics[bb=62bp 551bp 294bp 738bp]{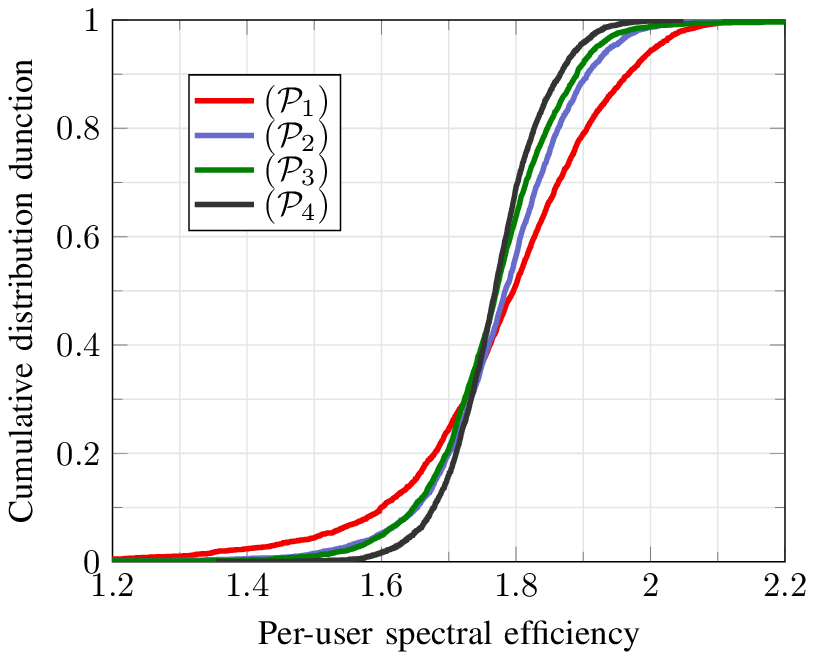}\label{fig:CDFK500}}
	\caption{CDF of per-user spectral efficiency for $(\mathcal{P}_{1})$-$(\mathcal{P}_{4})$ for small scale and large-scale problems.}
	\label{fig:CDF}
	\end{centering}
\end{figure}
Next we plot the cumulative distribution function (CDF) of the per-user
SE obtained by the four considered utility functions. Two settings
are examined: $M=100$, $K=20$, $T_{c}=200$ symbols and $T_{p}=20$
symbols (cf. Fig. \ref{fig:CDFK20}), and $M=500$, $K=2000$, $T_{c}=1000$
and $T_{p}=200$ (cf. Fig. \ref{fig:CDFK500}). We can observe in
Fig. \ref{fig:CDF} that the median values of the total achieved SEs are more
or less the same for all the four utility functions. The $95\%$-likely
achievable downlink SE is decreasing from $(\mathcal{P}_{1})$ to
$(\mathcal{P}_{4})$ which can be explained by the fact that the following
inequality holds: $\mathrm{SE}^{(\mathcal{P}_{1})}>\mathrm{SE}^{(\mathcal{P}_{2})}>\mathrm{SE}^{(\mathcal{P}_{3})}>\mathrm{SE}^{(\mathcal{P}_{4})}$,
where $\mathrm{SE}^{(\mathcal{P}_{i})}$ denotes the per-user spectral efficiency obtained by
solving $(\mathcal{P}_{i})$ \cite{Luo:SpectrumManagement:2008}.
It is also known that the order is reversed in terms of fairness.
Consequently, the CDF of the per-user SE of $(\mathcal{P}_{4})$ (i.e.
the MRmax problem) has the steepest slope and that of the SEmax problem
is more spread. It is particularly interesting to see that the difference
on the CDF of the per-user SE of all four utility metrics is marginal
for large-scale cell-free massive MIMO. This simulation result again
confirms that cell-free massive MIMO can deliver universally good
services to all users in the system.%
\begin{comment}
Figure~\ref{fig:CDFK20-1} is similar to Figure~\ref{fig:CDF} so
it can be skipped.
\end{comment}

\begin{figure}[tbh]
	\begin{centering}
		\textsf{\includegraphics[bb=62bp 550bp 267bp 738bp]{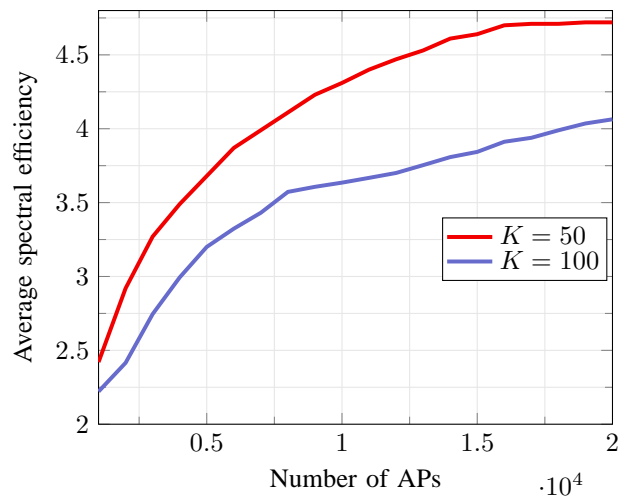}}
		\par\end{centering}
	\caption{Average spectral efficiency versus number of APs for $K=100$ and
		$K=50$ users for $D=1$.}
	\label{fig:avgSEvsAPs}
\end{figure}
In the next experiment, we investigate how the average SE varies as
a function of the number of APs. In particular, Fig. \ref{fig:avgSEvsAPs}
shows the average spectral efficiency as a function of the number
of APs for $K=100$ and $K=50$ users in an area of $D=1$. We can
see that the average SE increases quickly when the number of APs is
less 1000 for both $K=50$ and $K=100$ users, and it starts to saturate
when the number of APs is larger. The reason is that for a given user,
there should be a certain number of APs (i.e closest APs) that truly
provides macro diversity gains to that user in terms of SE. Thus,
a user may be served by a subset of APs to achieve similar SE performance.
This can help reduce the overhead in cell-free massive MIMO. Further
insights into this are discussed in the next numerical experiment.

\begin{figure}[tbh]
	\begin{centering}	
	\includegraphics[bb=62bp 556bp 288bp 741bp]{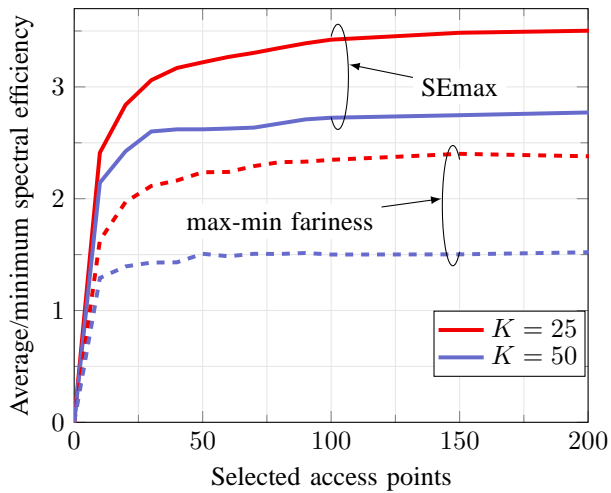}\caption{Average spectral efficiency versus number of selected APs for for
		$M=500$ and $D=1$.}
	\label{fig:usercentric}
	\end{centering}
\end{figure}
In Fig. \ref{fig:usercentric}, we consider a scenario with $M=500$
APs and study how many APs are effectively required for each user.
In particular, we plot the average spectral efficiency as a function
of the number of selected APs per user for two utility functions:
SEmax and MRmax. For a given user, a number of APs is simply selected
based on their large-scale fading coefficients. From Fig. \ref{fig:usercentric},
we can observe that not all 500 APs are needed to serve 25 or 50 users
in an area of \SI{1}{\km\squared}. Instead a smaller number of APs
per user can can yield nearly the same performance. For example, we
only need to assign less than $100$ APs to a user to achieve $95\%$ of the
SE of the full system.

\begin{figure}[tbh]
	\begin{centering}
		\includegraphics[bb=62bp 550bp 286bp 738bp]{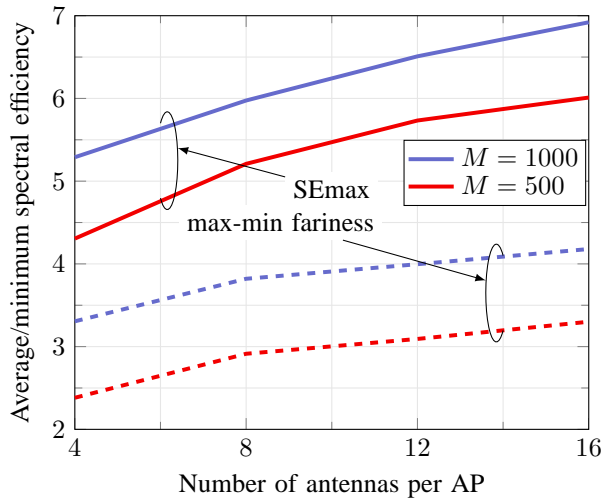}
		\par\end{centering}
	\caption{Average spectral efficiency with respect to the number of antenna
		at each AP for $K=50,D=1.$}
	\label{fig:SEvsTx}
\end{figure}
Finally, we investigate the effect of increasing the number of antennas
per AP on the sum SE and minimum SE. Specifically, we plot
the average SE and minimum SE with respect to the number of antennas
for both $M=500$ and $M=1000$ APs. The number of users is fixed
to $K=50$. Expectedly, the SE increases with the increase in the
number of antennas per APs but the increase tends to be small when
the number of antennas is sufficiently large. The reason is that for
a large number of APs channel harderning and favorable propagation
can be achieved by a few antennas per AP. Specifically, we can see
that the SE for the case of $1000$ APs and $4$ antennas per AP is
larger than the SE for the case of 500 APs and 8 antennas per AP.
Therefore, for large-scale cell-free massive MIMO, having more APs with a few antennas each seems to be more beneficial than having
fewer APs with more antennas each.

\section{Conclusion}

We have considered the downlink of cell-free massive MIMO and aimed
to maximize four utility functions, subject to a  power constraint
at each AP. Conjugate beamforming has been adopted, resulting in a
power control problem for which an accelerated project gradient method
has been proposed. Particularly, the proposed solutions only requires
the first order information of the objective and, in particular, each
iteration of the proposed solutions can be computed by closed-form
expressions. We have numerically shown that the proposed methods can
achieve (nearly) the same SE as a known SCA-based method but with much lesser
run time. For the first time, we have evaluated the SE performance
of cell-free massive MIMO for an area of \SI{10}{\km\squared}, consisting
of up to \num{10000} APs, whereby the achieved SE can be up to 200
(bit/s/Hz). We have also found that the SE performance of the four
utility functions is quite similar for large-scale cell-free massive
MIMO, confirming again that cell-free massive MIMO can provide uniformed
services to all users.

\appendix{}

\subsection{\label{apx:projection}Solution to the Projection onto $\mathcal{S}$}

The solution of \eqref{eq:project:subprob} can be found using the
KKT conditions given by \begin{subequations}
	\begin{align}
	\nabla_{\boldsymbol{\mu}_{m}}\mathcal{L}=2(\boldsymbol{\mu}_{m}-\mathbf{x}_{m})+2\lambda\boldsymbol{\mu}_{m} & =0,\label{eq:gradlangrangian}\\
	\lambda(||\boldsymbol{\mu}_{m}||^{2}-\frac{1}{N}) & =0,\label{eq:compslack}\\
	||\boldsymbol{\mu}_{m}||^{2} & \leq\frac{1}{N},\label{eq:normball}\\
	\lambda & \geq0.
	\end{align}
\end{subequations}Applying the constraint $\boldsymbol{\mu}_{m}\geq0$
to \eqref{eq:gradlangrangian}, we get 
\begin{equation}
\mathbf{x}_{m}=(1+\lambda)\boldsymbol{\mu}_{m}\geq0.\label{eq:positivemu}
\end{equation}
If $\lambda=0$, the stationary condition in \eqref{eq:gradlangrangian}
results in
\begin{equation}
\boldsymbol{\mu}_{m}=\mathbf{x}_{m},\label{eq:lambdazero}
\end{equation}
and \eqref{eq:normball} gives $\boldsymbol{\mu}_{m}=\mathbf{x}_{m}\leq\frac{1}{N}$,
which corresponds to the case where $\mathbf{x}_{m}$ lies in $\mathcal{S}$.
If $\lambda>0$, the complementary slackness in \eqref{eq:compslack}
implies that $||\boldsymbol{\mu}_{m}||^{2}=\frac{1}{N}$. The equality
in \eqref{eq:positivemu} can further be written as
\begin{align}
\boldsymbol{\mu}_{m}\trans\mathbf{x}_{m} & =(1+\lambda)\boldsymbol{\mu}_{m}\trans\boldsymbol{\mu}_{m}=(1+\lambda)\frac{1}{N}.
\end{align}
Since $\lambda>0$, we get $N\boldsymbol{\mu}_{m}\trans\mathbf{x}_{m}-1>0$
or $\boldsymbol{\mu}_{m}\trans\mathbf{x}_{m}>\frac{1}{N}$. By using
the Cauchy\textendash Schwartz inequality, we can write $||\boldsymbol{\mu}_{m}||\,||\mathbf{x}_{m}||>\frac{1}{N}$
or $||\mathbf{x}_{m}||^{2}>\frac{1}{N}$, which refers to the case
where $\mathbf{x}_{m}$ lies outside $\mathcal{S}$. Furthermore,
substituting $\lambda=N\boldsymbol{\mu}_{m}\trans\mathbf{x}_{m}-1$
into \eqref{eq:positivemu}, we get 
\begin{equation}
\mathbf{x}_{m}=N(\boldsymbol{\mu}_{m}\trans\mathbf{x}_{m})\boldsymbol{\mu}_{m},\label{eq:parallelx_mu}
\end{equation}
which means that $\boldsymbol{\mu}_{m}$ is parallel to $\mathbf{x}_{m}$
or $\boldsymbol{\mu}_{m}=a\mathbf{x}_{m}$ where $a$ is a constant.
Next using \eqref{eq:parallelx_mu} gives $a=\frac{1/\sqrt{N}}{||\mathbf{x}_{m}||^{2}}$
and therefore 
\begin{equation}
\boldsymbol{\mu}_{m}=\frac{1/\sqrt{N}}{||\mathbf{x}_{m}||^{2}}\mathbf{x}_{m}.\label{eq:lambdanonzero}
\end{equation}
Combining \eqref{eq:positivemu}, \eqref{eq:lambdazero} and \eqref{eq:lambdanonzero}
results in
\begin{equation}
\mu_{m}=\frac{\sqrt{1/N}}{\max\left(\sqrt{1/N},||[\mathbf{x}_{m}]_{+}||\right)}[\mathbf{x}_{m}]_{+},
\end{equation}
which completes the proof.

\subsection{\label{apx:LipschitzConstant}Lipschitz Constant of $\nabla f(\boldsymbol{\mu})$}

Assessing the Lipschitz constant of $\nabla f(\boldsymbol{\mu})$
for the four problems discussed in the paper boils down to finding
the Lipschitz constant of $\nabla\mathrm{SE}_{k}(\boldsymbol{\mu})$.
A convenient way to do this is to rewrite the function $\mathrm{SE}_{k}(\boldsymbol{\mu})$
in the form of single variable $\boldsymbol{\mu}$ which can be done
by denoting $\mathbf{A}_{i}\triangleq\mathbf{I}_{M}\otimes\mathbf{e}_{i}\trans$.
Then, we can write $\bar{\boldsymbol{\mu}}_{i}$ as $\mathbf{A}_{i}\boldsymbol{\mu}$,
and thus $\gamma_{k}(\boldsymbol{\mu})$, the SINR of the $k$-th
user, can be rewritten as
\begin{equation}
\gamma_{k}(\boldsymbol{\mu})=\frac{\overbrace{\zeta_{d}(\bar{\boldsymbol{\nu}}_{kk}\trans\mathbf{A}_{k}\boldsymbol{\mu})^{2}}^{b_{k}(\boldsymbol{\mu})}}{\underbrace{\zeta_{d}\left(\sum_{i\neq k}^{K}(\bar{\boldsymbol{\nu}}_{ik}\trans\mathbf{A}_{i}\boldsymbol{\mu})^{2}+\frac{1}{N}\sum_{i=1}^{K}||\bar{\mathbf{D}}_{k}\mathbf{A}_{i}\boldsymbol{\mu}||_{2}^{2}\right)}_{c_{k}(\boldsymbol{\mu})}+\frac{1}{N^{2}}}.
\end{equation}
The gradient of $\mathrm{SE}_{k}(\boldsymbol{\mu})$ with respect
to $\boldsymbol{\mu}$ is found as 
\begin{equation}
\nabla\mathrm{SE}_{k}(\boldsymbol{\mu})=\nabla_{\boldsymbol{\mu}}\log(1+\gamma_{k}(\boldsymbol{\mu}))
=\frac{\nabla_{\boldsymbol{\mu}}\left(b_{k}(\boldsymbol{\mu})+c_{k}(\boldsymbol{\mu})\right)}{b_{k}(\boldsymbol{\mu})+c_{k}(\boldsymbol{\mu})+\frac{1}{N^{2}}}-\frac{\nabla_{\boldsymbol{\mu}}c_{k}(\boldsymbol{\mu})}{c_{k}(\boldsymbol{\mu})+\frac{1}{N^{2}}},\label{eq:gradGamma-1}
\end{equation}
where $\nabla_{\boldsymbol{\mu}}b(\boldsymbol{\mu})$ and $\nabla_{\boldsymbol{\mu}}c(\boldsymbol{\mu})$
are given by 
\begin{equation}
\nabla_{\boldsymbol{\mu}}b_{k}(\boldsymbol{\mu})=2\zeta_{d}\mathbf{A}_{k}\trans\bar{\boldsymbol{\nu}}_{kk}\bar{\boldsymbol{\nu}}_{kk}\trans\mathbf{A}_{k}\boldsymbol{\mu},\label{eq:BkGradient-1}
\end{equation}

\begin{equation}
\nabla_{\boldsymbol{\mu}}c_{k}(\boldsymbol{\mu})=2\zeta_{d}\left(\sum_{i\neq k}^{K}\mathbf{A}_{i}\trans\bar{\boldsymbol{\nu}}_{ik}\bar{\boldsymbol{\nu}}_{ik}\trans\mathbf{A}_{i}+\frac{1}{N}\sum_{i=1}^{K}\mathbf{A}_{i}\trans\bar{\mathbf{D}}_{k}^{2}\mathbf{A}_{i}\right)\boldsymbol{\mu}.\label{eq:CkGradient-1}
\end{equation}
Now the Lipschitz continuity of $\nabla_{\boldsymbol{\mu}}b_{k}(\boldsymbol{\mu})$
and $\nabla_{\boldsymbol{\mu}}c_{k}(\boldsymbol{\mu})$ is obvious,
and so is that of $\nabla\mathrm{SE}_{k}(\boldsymbol{\mu})$. Although
we can compute the Lipschitz constant of $\nabla\mathrm{SE}_{k}(\boldsymbol{\mu})$,
this is quite involved and not necessary since a line search is used
to find a proper step size.
\vspace{-0.5cm}
\subsection{Convergence Proof of Algorithm \ref{alg:mAPG} \label{sec:Convergence-Proof}}

The proof is due to \cite{Li2015}. First, we note that since $\nabla f(\boldsymbol{\mu})$
is Lipschitz continuous and a line search is used to find a proper
step size in Algorithm \ref{alg:mAPG_LS}, it is sufficient to prove
the convergence Algorithm \ref{alg:mAPG}. We begin with by recalling
an important inequality of a $L$-smooth function. Specifically, for
a function $f(\mathbf{x})$ has the Lipschitz continuous gradient with a constant
$L_{f}$, the following inequality holds
\begin{equation}
f(\mathbf{y})\geq f(\mathbf{x})+\bigl\langle\nabla f\bigl(\mathbf{x}\bigr),\mathbf{y}-\mathbf{x}\bigr\rangle-\frac{L_{f}}{2}||\mathbf{y}-\mathbf{x}||^{2}.\label{eq:Lips:grad:inequality}
\end{equation}
The projection in Step \ref{gradmu} of Algorithm \ref{alg:mAPG}
can be written as
\begin{equation}
\mathbf{v}^{n+1}=\underset{\boldsymbol{\mu}\in\mathcal{S}}{\arg\min}\bigl\Vert\boldsymbol{\mu}-\boldsymbol{\mu}^{n}-\alpha\nabla f\bigl(\boldsymbol{\mu}^{n}\bigr)\bigr\Vert^{2}
=\underset{\boldsymbol{\mu}\in\mathcal{S}}{\arg\max}\ \bigl\langle\nabla f\bigl(\boldsymbol{\mu}^{n}\bigr),\boldsymbol{\mu}-\boldsymbol{\mu}^{n}\bigr\rangle-\frac{1}{2\alpha}||\boldsymbol{\mu}-\boldsymbol{\mu}^{n}||^{2},\label{eq:theta:project:rewrite}
\end{equation}
where we have used the fact that $||\mathbf{a}-\mathbf{b}||^{2}=||\mathbf{a}||^{2}+||\mathbf{b}||^{2}+2\left\langle \mathbf{a},\mathbf{b}\right\rangle $.
Note that when $\boldsymbol{\mu}=\boldsymbol{\mu}^{n}$, the objective
in the above problem is $0$, and thus we have 
\begin{equation}
\bigl\langle\nabla f\bigl(\boldsymbol{\mu}^{n}\bigr),\mathbf{v}^{n+1}-\boldsymbol{\mu}^{n}\bigr\rangle-\frac{1}{2\alpha}||\mathbf{v}^{n+1}-\boldsymbol{\mu}^{n}||^{2}\geq0.
\end{equation}
Applying \eqref{eq:Lips:grad:inequality} yields
\begin{align}
f(\mathbf{v}^{n+1}) & \geq f\bigl(\boldsymbol{\mu}^{n}\bigr)+\bigl\langle\nabla f\bigl(\boldsymbol{\mu}^{n}\bigr),\mathbf{v}^{n+1}-\boldsymbol{\mu}^{n}\bigr\rangle
-\frac{L_{f}}{2}\bigl\Vert\mathbf{v}^{n+1}-\boldsymbol{\mu}^{n}\bigr\Vert^{2}\nonumber \\
& \geq f\bigl(\boldsymbol{\mu}^{n}\bigr)+\bigl(\frac{1}{2\alpha}-\frac{L_{f}}{2}\bigr)\bigl(\bigl\Vert\mathbf{v}^{n+1}-\boldsymbol{\mu}^{n}\bigr\Vert^{2}.\label{eq:theta:project:ineq}
\end{align}
It is easy to see that $f(\mathbf{v}^{n+1})\geq f\bigl(\boldsymbol{\mu}^{n}\bigr)$
if $\alpha<\frac{1}{L_{f}}$. From Step \ref{alg:mAPG:update}, if
$f(\mathbf{z}^{n+1})\geq f(\mathbf{v}^{n+1})$, then
\begin{align}
\boldsymbol{\mu}^{n+1} & =\mathbf{z}^{n+1},f\bigl(\boldsymbol{\mu}^{n+1}\bigr)=f(\mathbf{z}^{n+1})\geq f(\mathbf{v}^{n+1}).\label{eq:extra:ineq}
\end{align}
Similar if $f(\mathbf{z}^{n+1})<f(\mathbf{v}^{n+1})$, then 
\begin{align}
\boldsymbol{\mu}^{n+1} & =\mathbf{v}^{n+1},f\bigl(\boldsymbol{\mu}^{n+1}\bigr)=f(\mathbf{v}^{n+1}).\label{eq:theta:project:update}
\end{align}
From \eqref{eq:theta:project:ineq}, \eqref{eq:extra:ineq}, and \eqref{eq:theta:project:update}
we have 
\begin{equation}
f\bigl(\boldsymbol{\mu}^{n+1}\bigr)\geq f(\mathbf{v}^{n+1})\geq f\bigl(\boldsymbol{\mu}^{n}\bigr).
\end{equation}
Since the feasible set of the considered problems is compact convex,
the iterates $\{\mathbf{v}^{n}\}$ and $\{\boldsymbol{\mu}^{n}\}$
are both bounded and thus $\{\boldsymbol{\mu}^{n}\}$ has accumulation
points. As shown above, $f\bigl(\boldsymbol{\mu}^{n}\bigr)$ is nondecreasing,
$f$ has the same value, denoted by $f^{\ast}$, at all the accumulation
points. From \eqref{eq:theta:project:ineq}, we have
\begin{equation}
f\bigl(\boldsymbol{\mu}^{n+1}\bigr)-f\bigl(\boldsymbol{\mu}^{n}\bigr)\geq f(\mathbf{v}^{n+1})-f\bigl(\boldsymbol{\mu}^{n}\bigr)
\geq\bigl(\frac{1}{2\alpha}-\frac{L_{f}}{2}\bigr)\bigl(\bigl\Vert\mathbf{v}^{n+1}-\boldsymbol{\mu}^{n}\bigr\Vert^{2},
\end{equation}
which results in
\begin{equation}
\infty>f^{\ast}-f\bigl(\boldsymbol{\mu}^{1}\bigr)\geq\sum_{n=1}^{\infty}\bigl(\frac{1}{2\alpha}-\frac{L_{f}}{2}\bigr)\bigl(\bigl\Vert\mathbf{v}^{n+1}-\boldsymbol{\mu}^{n}\bigr\Vert^{2}.
\end{equation}
Since $\alpha<\frac{1}{L_{f}}$, we can conclude that 
\begin{equation}
\bigl\Vert\mathbf{v}^{n+1}-\boldsymbol{\mu}^{n}\bigr\Vert\to0\ \textrm{as}\ n\to\infty.\label{eq:iterate:converge}
\end{equation}
The convergence proof of Algorithm \ref{alg:mAPG} to a critical point
of $(\mathcal{P}_{1})$ follows the same arguments as those in \cite{Li2015}
and thus, is omitted for the sake of brevity.
%\vspace{-0.5cm}
\bibliographystyle{IEEEtran}
\bibliography{IEEEabrv,bibTex,references}

\end{document}